\documentclass[twocolumn,preprintnumbers,nofootinbib,prd,superscriptaddress,aps]{revtex4-1}

\usepackage[utf8]{inputenc}
\usepackage{graphicx,amssymb,amsmath,amsthm,amsfonts,epstopdf,epsfig,epsf,times}
\usepackage[linktocpage]{hyperref}
\usepackage[usenames]{color}
\usepackage{epstopdf}

\graphicspath{{plots/}} 

\usepackage{bm}
\usepackage{dcolumn}
\usepackage{latexsym}
\usepackage{rotating}
\usepackage{hyperref}
\usepackage{color}
\usepackage{longtable}
\usepackage{enumerate}
\usepackage{tensor}
\usepackage{stmaryrd}
\usepackage[normalem]{ulem}
\usepackage{mathtools}
\usepackage{url}
\definecolor{coolblack}{rgb}{0.0, 0.18, 0.39}
\definecolor{darkred}{rgb}{0.5,0,0}
\definecolor{darkgreen}{rgb}{0,0.5,0}
\definecolor{darkblue}{rgb}{0,0,0.5}
\definecolor{lapislazuli}{rgb}{0.15, 0.38, 0.61}
\definecolor{venetianred}{rgb}{0.78, 0.03, 0.08}
\definecolor{bleudefrance}{rgb}{0.19, 0.55, 0.91}
\definecolor{dogwoodrose}{rgb}{0.84, 0.09, 0.41}
\definecolor{dogwoodrose}{rgb}{0.84, 0.09, 0.41}
\definecolor{darkorgane}{rgb}{1,0.549,0}
\hypersetup{colorlinks=true, citecolor=darkblue, linkcolor=darkblue,
urlcolor = darkblue}

\definecolor{olive}{rgb}{0.5, 0.5, 0.0}

\newcommand{\ben}{\begin{enumerate}}
\newcommand{\een}{\end{enumerate}}

\newcommand{\leqsim}{\,\mbox{{\scriptsize $\stackrel{<}{\sim}$}}\,}
\newcommand{\geqsim}{\,\mbox{{\scriptsize $\stackrel{>}{\sim}$}}\,}

\def\be{\begin{equation}}
\def\ee{\end{equation}}
\newcommand{\beq}{\begin{eqnarray}}
\newcommand{\eeq}{\end{eqnarray}}
\newcommand{\ba}{\begin{align}}
\newcommand{\ea}{\end{align}}

\definecolor{darkorange}{rgb}{1,0.549,0}

\definecolor{tangerineyellow}{rgb}{1.0, 0.8, 0.0}
\definecolor{springgreen}{rgb}{0.0, 1.0, 0.5}
\definecolor{bostonuniversityred}{rgb}{0.8, 0.0, 0.0}

\def\be{\begin{equation}}
\def\ee{\end{equation}}
    
\newcommand{\bea}{\begin{eqnarray}}
\newcommand{\eea}{\end{eqnarray}}
\def\l{\left}
\def\r{\right}

\begin{document}\title {\large Soliton boson stars, Q-balls and the causal Buchdahl bound}

\author{Mateja Bošković}
\affiliation{SISSA, Via Bonomea 265, 34136 Trieste, Italy and INFN Sezione di Trieste}
\affiliation{IFPU - Institute for Fundamental Physics of the Universe, Via Beirut 2, 34014 Trieste, Italy}
\author{Enrico Barausse}
\affiliation{SISSA, Via Bonomea 265, 34136 Trieste, Italy and INFN Sezione di Trieste}
\affiliation{IFPU - Institute for Fundamental Physics of the Universe, Via Beirut 2, 34014 Trieste, Italy}

\begin{abstract}
Self-gravitating non-topological solitons whose potential admits multiple vacua are promising candidates for exotic compact objects. Such objects can arise in several extensions of the Standard Model and could be produced in the early Universe.
In this work, we focus on objects made from complex scalars (gravitating Q-balls/soliton boson stars),
deriving analytic solutions in spherical symmetry and comparing them with  fully numerical ones.
In the high-compactness limit we find that these objects present an effectively linear equation of state, thus  saturating the Buchdahl limit with the causality constraint.
Far from that limit, these objects behave either as flat space-time Q-balls or (in the low-compactness limit) as mini boson stars stabilized by quantum pressure. We establish the robustness of this picture by analyzing a variety of potentials (including cosine, quartic and sextic ones). 
\end{abstract}

\date{\today}

\maketitle

\tableofcontents

\section{Introduction}\label{sec:intro}


Gravitational wave (GW)~\cite{Taylor:1982zz,LIGOScientific:2016aoc,LIGOScientific:2017vwq,LIGOScientific:2019fpa,LIGOScientific:2020ibl,LIGOScientific:2021qlt}  and
electromagnetic (EM)~\cite{PhysRevLett.9.439,1965Sci...147..394B,Reid:1999kt,Ghez:2000ay,EventHorizonTelescope:2019dse,EventHorizonTelescope:2019ggy} observations provide a unique opportunity to understand the formation, dynamics and environment of compact objects in the Universe, as well as to test
gravity in the hitherto unexplored strong-field regime.
Besides known compact objects, such as white dwarfs, neutron  stars and black holes (BHs),  
more exotic  compact objects (ECOs) \cite{Giudice:2016zpa,Cardoso:2019rvt} may also exist
as a result of extensions of the Standard Model,
and might
 represent a fraction of the dark matter (DM) or be relics from  processes in the early Universe \cite{Bertone:2019irm}. In particular, ECOs with compactness $\mathcal{C} \equiv G M/(R c^2) > 1/3$ ($M$ and $R$ being respectively the mass and radius of the body),
 also referred to as ``ultra-compact objects'', may be hard to distinguish from BHs if they present a light ring~\cite{Cardoso:2019rvt}. However, toy models for ECOs (such as wormholes \cite{Visser:1995cc},  gravastars \cite{Mazur:2001fv}, anisotropic fluid stars \cite{Raposo:2018rjn} etc) have  point at a rich range of possible observational features distinguishing these
objects from BHs, e.g.
non-vanishing Love numbers \cite{Cardoso:2017cfl}, distinct post-merger phase for binary systems \cite{Toubiana:2020lzd}, GW echoes \cite{Barausse:2014tra,Cardoso:2016rao, Cardoso:2016oxy, Cardoso:2019rvt} etc.

Many ECO toy models, however, are not realistic candidates, as they contradict some of the following reasonable requirements:
stability (on relevant astrophysical/cosmological scales); existence of possible production mechanisms and astrophysical/cosmological formation channels; consistency with known and tested physics; and embedding in a plausible beyond-Standard Model theory. A more promising way to construct realistic ECO models is to consider solitons, i.e. localized, finite-energy and stable solutions of the equations of motion of a field theory. Derrick's theorem (e.g.~\cite{Nastase:2019pzh}) is a powerful constraint on the existence of these solutions, as it prohibits static non-trivial solutions for a set of real scalar fields in $D \geq 2$, where $D$ is the number of space dimensions. In order to evade Derrick's theorem, one must either consider topologically non-trivial configurations (topological solitons) or consider a theory with conserved charge (non-topological solitons)~\cite{Lee:1991ax,Nastase:2019pzh}.  In this work, we will consider the latter type of solitons\footnote{Examples of topological solitons in the strong gravity context are cosmic strings \cite{Helfer:2018qgv} and sine-Gordon stars \cite{Franzin:2018pwz}.}.


The simplest examples of non-topological solitons are boson stars, i.e.  configurations made from complex scalars $\Phi$ minimally coupled to gravity
and admitting a $U(1)$ symmetry. Boson stars are regular horizonless solutions in general relativity. These objects have been thoroughly studied as potential ECOs~\cite{Jetzer:1991jr,Schunck:2003kk,Liebling:2012fv, Cardoso:2019rvt, Visinelli:2021uve}. Among the different models of boson stars, the  most investigated are mini-boson stars \cite{Kaup:1968zz} (MBSs; which only have a mass term in the potential), self-interacting boson stars \cite{Colpi:1986ye} (SIBSs; whose potential also includes a  repulsive $\lambda \Phi^4$ term) and solitonic boson stars \cite{Friedberg:1986tq} (SBSs), which present the potential
\begin{align}\label{eq:V6qball}
V = \mu^2 |\Phi|^2 \Big(1 -2 \frac{|\Phi|^2}{\sigma_0^2} \Big)^2 \,,
\end{align}
where $\sigma_0$ is the degenerate vacuum and $\mu$ is a scalar mass\footnote{More complicated examples of boson stars   are either made from multiple \cite{Bezares:2018qwa} or higher-spin fields \cite{Brito:2015pxa},  have a bigger~\cite{Herdeiro:2018djx} or gauged symmetry group~\cite{Liebling:2012fv}, or include both bosonic and fermionic components~\cite{Henriques:1989ar}.}. 

These models describe three distinct physical mechanisms of stabilizing the configurations (see Sec. \ref{sec:oom}).
In MBSs/SIBSs, that mechanism is provided by the equilibrium between gravity and respectively quantum pressure/repulsive pressure. For SBSs, which have a bubble-like structure in the most compact part of the parameter space, we will see that
accumulation of the energy near the surface gives rise to a surface tension between different vacua,
which allows for
developing massive and highly compact configurations. Because of this, SBSs are the only model
among the aforementioned three
that exists also in the $G \to 0$ $(M_{\rm Pl} \to \infty)$ limit. (In that limit, SBSs are also referred to as Q-balls \cite{Coleman:1985ki}).
The three models mentioned above should be considered as an illustration of the different mechanisms that can bind  boson stars together. Models
more motivated from the particle physics point of view
would follow from renormalizable~\cite{Freitas:2021cfi} or effective field theories for the scalar~\cite{Krippendorf:2018tei, Barranco:2021auj}. However, we expect that the macroscopic behaviour of the
resulting boson stars should still fall, at least qualitatively, in one of the three classes mentioned above,
as we will argue here in the case of SBSs.

Boson stars generally satisfy the soundness checks that we  mentioned above. Non-rotating boson stars have at least one stable branch in the low-compactness (MBS) limit \cite{Seidel:1990jh, Liebling:2012fv}. Although rotation can lead to instabilities~\cite{DiGiovanni:2020ror}, it has been  recently shown that sufficiently strong self-interactions   can sustain it~\cite{Siemonsen:2020hcg, Dmitriev:2021utv}.  As these objects are constructed from
covariant Lorentz-invariant actions, they are consistent with causality  from the start (something often not clear with ad hoc ECO models), and the modelling of their dynamics is in principle accessible. Boson stars can form from   scalar collapse, with the excess field `evaporating' (gravitational cooling) \cite{Seidel:1993zk}. Once a population of non-rotating boson stars is formed, it can reproduce in binaries through the channel $\rm{BS}+\rm{BS} \to \rm{BS}$ \cite{Palenzuela:2017kcg}. In this work we will focus on SBSs, as they stand out as the most compact ones, with $\mathcal{C}_{\rm max} \approx 0.33$ \cite{Cardoso:2016oxy} or even $\mathcal{C}_{\rm max} \approx 0.348$ \cite{Friedberg:1986tq}.


  SBSs are the simplest representatives of non-topological solitons that exist in the $M_{\rm Pl} \to \infty$ limit and admit a false or degenerate vacuum in one of the bosonic degrees of freedom\footnote{The study of these objects was initiated by T. D. Lee and  collaborators in the 1980s \cite{Lee:1986ts,Friedberg:1986tp,Friedberg:1986tq,Lee:1986tr,Lee:1987rr,Lee:1991ax}.}. Thus forth, we will focus on this specific model, in order to get an insight on the general features of the whole class of objects and
potentially on those of
more realistic and complicated models that we plan to address in the future \cite{Lee:1986tr,Bahcall:1989ff,Hong:2020est,Gross:2021qgx,FSS}.  

Besides simplicity, there  are also at least two additional reasons to be interested in this kind of objects.
First, Q-balls/SBSs  can be formed in the early Universe, through scalar fragmentation after inflation~\cite{Cotner:2019ykd,Amin:2019ums}, thermal phase transitions~\cite{Krylov:2013qe} or solitosynthesis~\cite{Postma:2001ea,Croon:2019rqu}. Therefore, the possible observation of SBSs or similar ECOs can shed light on beyond-Standard Model physics,
 from baryogenesis to dark matter~\cite{Rubakov:2017xzr}. The charge of these objects could be protected by an approximate, low-energy global symmetry \cite{Krippendorf:2018tei}, and depending on the specific model  a fraction of
the compact-object population of the universe could be in this form~\cite{Troitsky:2015mda}. Note that
 even if  SBSs are low-compactness at  formation, their subsequent interaction and merger could lead to the formation of more compact  configurations.

SBSs could also be  considered as a proxy for similar self-gravitating structures made from (real) bosons unprotected by a symmetry. Among the latter are axion stars \cite{Helfer:2017a}, but also oscillatons \cite{Nazari:2020fmk}, moduli stars \cite{Krippendorf:2018tei,Muia:2019coe} etc.  These objects are not strictly speaking solitons, and both the energy-momentum tensor and metric are time dependent \cite{Seidel:1991zh,Helfer:2017a,Boskovic:2018rub}. However, owing to the large number of particles, their decay is exponentially suppressed \cite{Kasuya:2002zs,Hertzberg:2010yz,Zhang:2020bec}.  Notwithstanding their pseudo-solitonic nature, these objects have  macroscopic properties qualitatively similar to the corresponding boson star models. For example, the $M$-$R$ curve of scalar stars
with a mass-term potential (oscillaton) differs only by (at most) a few percent
from the corresponding MBS model, and only in the most compact branch~\cite{Brito:2015yfh}. Axion stars and similar objects are expected to form in a wider variety of settings - through cosmological evolution of axion dark matter \cite{Marsh:2016rep,Hui:2021tkt,Schive:2014dra,Levkov:2018kau,Widdicombe:2018oeo,Arvanitaki:2019rax}, as inflation relics \cite{Amin:2014eta, Niemeyer:2019gab,Arvanitaki:2019rax} or through  other early Universe processes similar to those giving rise to complex configurations \cite{Krippendorf:2018tei,Cotner:2019ykd,Amin:2019ums}. Similar objects can also arise in certain scalar-tensor theories \cite{Barranco:2021auj}. Although an exact parallel between boson star models and pseudo-solitonic configurations warrants a more in-depth study, in particular when a false/degenerate vacuum exists, the static nature of  boson star spacetimes allows for
a technically easier preliminary analysis of these objects.


Besides the rich literature on Q-balls (starting from the seminal paper by Coleman~\cite{Coleman:1985ki}; see also~\cite{Nugaev:2019vru} for a recent review), the structure of relativistic SBSs has been investigated for  the simplest potential  \eqref{eq:V6qball}~\cite{Friedberg:1986tq,Kesden:2004qx,Macedo:2013jja,Cardoso:2016oxy,Siemonsen:2020hcg}, for more general sextic potentials~\cite{Kleihaus:2005me,Kleihaus:2011sx,Siemonsen:2020hcg} and for related potentials, such as the cosine one~\cite{Guerra:2019srj,Siemonsen:2020hcg}.  Quasi-normal modes and geodesics around SBSs have also been studied \cite{Macedo:2013jja},  their Love numbers have been calculated \cite{Cardoso:2017cfl,Sennett:2017etc}, and they have been simulated in  binary systems~\cite{Cardoso:2016oxy,Bezares:2017mzk,Palenzuela:2017kcg,Bezares:2018qwa,Helfer:2021brt,Bezares:2022obu}.

The relation between Q-balls and SBSs has been at least partially discussed previously~\cite{Multamaki:2002wk,Tamaki:2010zz,Tamaki:2011zza,Tamaki:2011bx} and the most dramatic effects of gravity are non-perturbative in $\sigma_0/M_{\rm Pl}$. In this regime, as $M_{\rm Pl} < \infty$, Q-ball limit is not realized. However, there is no major qualitative difference between Q-balls and SBSs in the part of the parameter space where the effects of gravity are not important and perturbative. For clarity, in the rest of the paper we will  refer to the  $M_{\rm Pl}\to \infty$ limit as Q-balls and to the gravitating case as SBSs.

Notwithstanding all these studies, to the best of our knowledge some basic aspects of SBSs  and non-topological solitons that admit degenerate vacua in general have not been appropriately addressed in the  literature. In this work we aim to fill this gap by constructing analytic descriptions of SBSs for the simplest potential \eqref{eq:V6qball} in the sub-Planck limit $\sigma_0/M_{\rm Pl} \ll 1$, by understanding the physics that sets the maximal compactness  of these objects, and by illuminating the non-trivial connection with the Q-ball limit.
We will establish the robustness and the limits of this picture by exploring the parameter space in $\sigma_0/M_{\rm Pl}$, and also by considering cosine, sextic and quartic potentials.
We will show that diverse choices of the  potential do not correspond to dramatically distinct macroscopic behaviours of SBSs,  and that the structure of  SBSs mostly depends on the distance between the central field and the scalar false vacuum, and on whether the false vacuum is sufficiently deep.

This paper is organized as follows. In Sec.~\ref{sec:qb} we will provide a brief review of Q-balls, upon which we build up description of SBS structure and properties
in Sec.~\ref{sec:gravitating}.
In Sec.~\ref{sec:par_space} we will explore the parameter space of these objects and the effect of the scalar potential on their structure.
The technical details of our numerical methods are presented in App. \ref{AppEKG}, while in App. \ref{app:rad} we discuss various definitions of the SBS radii and in App. \ref{app:analytic_uS} we provide some additional details on the analytic construction of SBSs.
Throughout this paper, we will employ a metric  signature $-+++$ and natural units $c=\hbar=1$, with $M^2_{\rm Pl}=1/8\pi G$ and $m^2_{\rm Pl}=1/G$. Note that in geometric units $(G=c=1)$  $M^2_{\rm Pl}/\hbar =1/8 \pi$ and $[\hbar]=[M]^2$.


%

\section{Q-balls: a review}\label{sec:qb}

The Lagrangian of a  scalar field with $U(1)$ symmetry in  flat space-time is given by
\begin{equation} \label{eq:qb_action}
\mathcal{L}_{\Phi}=-\partial_\mu \Phi^\dag \partial^\mu \Phi-V(|\Phi|)\,.
\end{equation}
The scalar field can be decomposed in Fourier modes as
\begin{equation} \label{eq:qb_field}
\Phi=\phi(r) e^{-i\omega t} \,,
\end{equation}
with time translations corresponding to a change in phase. Because of the $U(1)$ symmetry, and
as long as only one Fourier mode is excited, this
superficial time dependence does not propagate to any observable quantity, such as thermodynamics parameters: the density $\rho= - T^t_t $ , the radial pressure $P_{\rm rad}= T^r_r $ etc., and  Derrick's theorem is circumvented.

Varying the action   \eqref{eq:qb_action} with respect to $\phi$, one obtains the Klein-Gordon equation
\begin{eqnarray} \label{eq:qb_KG}
\phi''+\frac{ 2}{r} \phi' &=& - \frac{dU_\omega}{d\phi} \,, \\
U_\omega &=&   \frac{1}{2}( \omega^2 \phi^2 -V(\phi))  \,.
\end{eqnarray}
Note that this equation can be interpreted as the equation of motion of a Newtonian particle under a friction term and an effective potential $U_\omega $. The energy of the Newtonian particle is
\begin{eqnarray}
\mathcal{E}=\frac{1}{2} (\phi')^2 + U_\omega  \,,
\end{eqnarray}
and it is conserved only in $D=1$ dimensions, when the friction is absent.

In order for the field energy (i.e. the integrated energy density) $E=\int dV \rho$ to be finite, one must have $\phi(r \to \infty) \to 0$, which implies that the Q-ball is a localized object. Let us assume that at infinity the scalar is free, with leading mass term in the potential $V=\frac{1}{2}\mu^2\phi^2 + \mathcal{O}(\phi^4)$. As  $U_\omega \sim (\omega^2 - \mu^2) \phi^2 $\,,  $\phi \to 0$, we require $\omega < \mu$ for the scalar to converge to the vacuum state at  infinity. The leading order behaviour is of the form $\phi \sim \exp{(-\sqrt{\mu^2-\omega^2}r)}$, which implies zero ``velocity'' $\phi'$ at  infinity and $\mathcal{E}(\infty)=0$. The asymptotic Yukawa-like behaviour implies that the Q-ball radius is ill defined. We follow the common practice (e.g. \cite{Macedo:2013jja}) and arbitrarily define the radius as that enclosing 99$\%$ of the total mass $R_{99}$. (Note however that different conventions and definitions can be found in the literature, see App. \ref{app:rad}).

At the initial point (in the particle perspective), in order for the friction to be overcome one must have  $\phi'(r \to 0) \propto r^{1+\epsilon} \,,\epsilon>0$ i.e. $\phi'(0)=0$. As $U_\omega(0)=0$, the particle needs to be released at some point $\phi_c | U_\omega(\phi_c)>0 \,, U_\omega'(\phi_c)>0$ in order to overcome the friction and arrive at infinity with finite energy. The last condition implies that $U_\omega$ must admit an additional hill, or more formally there must be a non-trivial minimum of $V/\phi^2$ at some point $\phi_0 \neq 0$ and $\omega \geq \omega_0 \equiv \text{min}[V/\phi^2]$  \cite{Coleman:1985ki}. The simplest potential is thus of the form
\begin{align}
V = \mu^2 |\Phi|^2 \Big(1 - 2 \frac{|\Phi|^2}{\sigma_0^2} \Big)^2 \,,
\end{align}
with $\phi_0=\sigma_0/\sqrt{2}$. A theory with this potential is non-renormalizable, and thus must interpreted in the context of  effective or thermal field theories\footnote{If one considers non-Abelian groups, instead, renormalizable potentials can also admit stable Q-balls~\cite{Safian:1987pr}.}.

\begin{figure*}[th]
\includegraphics[width=\textwidth]{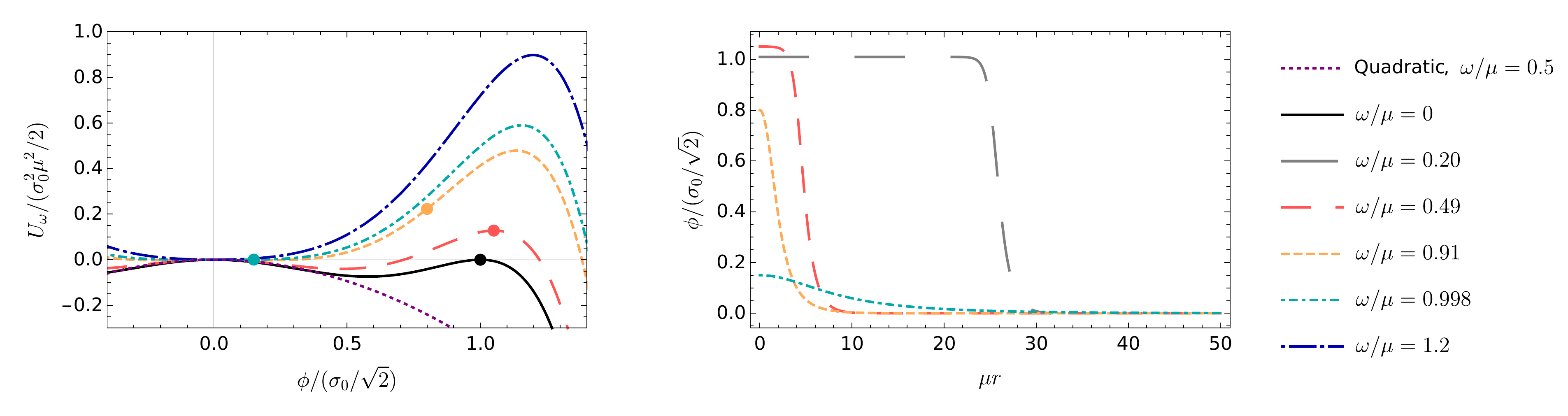}
\caption{(Left) Effective particle potential $U_\omega$ for various values of the field frequency $\omega$. The dots correspond to initial field values for physical configurations and all curves  represent the potential  \eqref{eq:V6qball}, except for the purple one which represents the quadratic potential $V=\mu^2 |\Phi|^2$ (that doesn't allow for solitonic solutions in flat space-time). (Right) Radial field profile (numerical results) for several physical configurations. Note  that the $\omega/\mu=0$ case corresponds to the infinitely large Q-ball with scalar $\phi=\sigma_0/\sqrt{2}$.}
    \label{fig:qb_pot_profile}
\end{figure*}

A sketch of the corresponding $U_\omega$ is shown in Fig. \ref{fig:qb_pot_profile}. In the limit $\omega=\omega_0$,
the fictitious particle is at the second hill $\phi_c \to \phi_0$, and the scalar would need an infinite amount of time to travel to the first hill (trivial vacuum). This scenario, in the Q-ball perspective, corresponds to an infinitely large configuration. The impact of the friction in this case can be neglected, as it is proportional to $1/R \to 0$. For a small but non-zero value of $\omega$ , the traveling time/Q-ball radius becomes finite,  but still large (thin-wall regime). In the opposite case when $\omega \to \mu$, the impact of friction is pronounced and the transition to the origin is smoother (thick-wall regime). The particle picture also demonstrates how one can construct these solutions numerically - through a shooting algorithm (see Sec. \ref{sec:grav_eq} and App. \ref{AppEKG} for the numerical formulation). If one releases the particle on the left of $\phi_c$ (e.g. somewhere around the valley for $\omega/\mu=0.5$), it will not have enough energy to reach the first hill. However, releasing the particle sufficiently near the second hill will let it reach the trivial vacuum with an excess of energy. In this way, one can iteratively find the true solution.

From the invariance under internal $U(1)$ symmetry, the Noether current and charge can be found as
\begin{eqnarray}
j^\mu &=& -i (\Phi^\dag \partial^\mu \Phi - \Phi \partial^\mu \Phi^\dag ) \,, \\
Q &\equiv& - \int dV j^t = 4 \pi \omega \int dr \,  r^2 \phi^2 \,.\label{eq:SBS_charge}
\end{eqnarray}
In order for the Q-ball not to decay to constituent scalars, the energy per unit charge should be lower than the particle mass:
\begin{eqnarray} \label{eq:qb_stab}
E < \mu Q \,.
\end{eqnarray}
In~\cite{Coleman:1985ki}, Coleman proved the stability of Q-balls, under the above restrictions on the form of the potential in classical field theory (necessary condition) and the requirement \eqref{eq:qb_stab} . Other aspects of  stability are discussed in \cite{Paccetti:2001uh,Sakai:2007ft}.

To obtain a rough understanding of the Q-ball properties, let us consider the potential \eqref{eq:V6qball}. When the central field is near the degenerate vacuum $\sigma_0$ (thin-wall regime), the main contribution to the mass of the object comes from the bulk $ \sigma^2_0 R^3$ and the surface tension $\delta^{-1} \sigma^2_0 R^2$, where $\delta \sim \mu^{-1}$ is the size of the potential wall. In equilibrium one has [with $(\Phi \sim \sigma_0 \exp{i\omega t})$]
\begin{equation} \label{eq:oom_SBS_R}
R \sim  \frac{\mu}{\omega^2}    \,.
\end{equation}
This rough argument provides the leading order behaviour of Q-balls in the thin-wall limit.

\subsection{Simplest Q-ball potential: analytic description} \label{sec:qb_simp}

We will now focus on the simplest potential with a global $U(1)$ symmetry and which admits stable Q-balls  \eqref{eq:V6qball}. For this potential, we can use the following scaling:
\begin{equation} \label{eq:qb_scale}
\mathsf{r} = \mu r \,,  w=\frac{\omega}{\mu}  \,, \varphi=\phi/(\sigma_0/\sqrt{2})\,,
\end{equation}
which makes the equations of motion dimensionless.

In \cite{Heeck:2020bau}, analytic profiles for Q-balls have been constructed (building on \cite{Paccetti:2001uh,Tsumagari:2008bv}) by matching solutions in three regimes -- interior (perturbative solution around $\varphi_c$), boundary (expansion around the radius) and  exterior (asymptotic). Here, we only mention these results, as we will partially review them (together with their curved space-time generalization) in the next section:
\begin{eqnarray}
\varphi_< &\approx& \varphi_+ \Big(1-c_< \frac{\sinh(\alpha \mathsf{r})}{\mathsf{r}} \Big) \,, \,  \label{eq:qb_interior} \\
\varphi_{B} &\approx&  \varphi_+  \frac{1}{\sqrt{1+2e^{2(\mathsf{r}-\mathsf{R^\ast})}}} \,,  \label{eq:qb_surface} \\
\varphi_{>} &\approx& \frac{ \varphi_+ c_>}{\mathsf{r}} e^{-\sqrt{1-w^2}\mathsf{r}} \label{eq:qb_asymp} \,,
\end{eqnarray}
where
\begin{eqnarray}
\alpha^2 = \frac{4}{3} \big( 1+3w^2+2\sqrt{1+3w^2} \big) \nonumber  \,,\\ \varphi^2_+=\frac{1}{3}(2+\sqrt{1+3w^2}) \label{eq:qb_central}\,.
\end{eqnarray}
Here, $\varphi_+$ corresponds to the maximum of  $U_\omega$. For $w \ll 1$, the maximum of $U_\omega$ lies close to the degenerate vacuum $\varphi \approx 1$; $\mathsf{R^\ast}=\mu R^\ast$ is the inflection point of the field $\varphi''(\mathsf{R^\ast})=0$, taken as the estimate of the size of the Q-ball. We confirm the observation from \cite{Heeck:2020bau} that the function \eqref{eq:qb_surface} does a good job at describing the numerical profile of the Q-ball, even outside its range of validity, in contrast with the interior and asymptotic approximants, which only work sufficiently close/far to the centre, respectively.

By matching  the fields~ \eqref{eq:qb_interior}-\eqref{eq:qb_asymp} and their derivatives at $\{ \mathsf{r_<}, \mathsf{r_>} \}$,  one finds~\cite{Heeck:2020bau}:
\begin{eqnarray}
\, c_< \approx \mathsf{R}^\ast e^{-2 \mathsf{R}^\ast} \label{eq:qb_c<}\,,
\, c_> \approx \sqrt{2} \mathsf{R^\ast} e^{\mathsf{R^\ast}} \,.
\end{eqnarray}
However, this matching is not sufficient to close the system. One needs also the energy balance condition
\begin{equation}
\mathcal{E}(\infty)-\mathcal{E}(0)=-\int^{\infty}_0 dr\big[ \frac{2}{r} (\phi')^2  \big] \,. \label{eq:qb_en_bal}
\end{equation}
In summary, we have five equations in five unknowns $\{c_<, c_>, \mathsf{R^\ast}, \mathsf{r_<}, \mathsf{r_>} \}$ as functions of $\omega$, or equivalently $\varphi_c(\omega)=\varphi_+ (1-c_<)$.
From Eq. \eqref{eq:qb_c<} we see that in the thin-wall regime (large Q-balls), the deviation of the field from the central value  in the interior zone is exponentially suppressed. In the outer zone, the field itself is exponentially suppressed, and thus the integral on the right hand side of  Eq. \eqref{eq:qb_en_bal} is dominated by
the boundary zone.

From  invariance under  time reversal, we expect  $R^\ast$ to be a Laurent polynomial in even powers of $\omega$. From  Eq. \eqref{eq:qb_en_bal} and introducing $z=\mathsf{r}-\mathsf{R^\ast}$, we then have
\begin{eqnarray} \label{eq:qb_en_bal_expl}
&& \left(e^{-2 \mathsf{R^\ast}} \mathsf{R^\ast}-1\right)^2 (2+\sqrt{1+3w^2}) \times \nonumber \\
&& \Big[w^2 \mathsf{R^\ast} -\mathsf{R^\ast}  \Big(1-\frac{1}{3} \big(e^{-2  \mathsf{R^\ast}}  \mathsf{R^\ast}-1\big)^2 (\sqrt{3 w^2+1}+2)\Big)^2 \Big]  \nonumber \\
&&
= 4 \varphi_+ \int^{+\infty}_{-\infty}\frac{dz}{1 + \frac{z}{\mathsf{R^\ast}}}\frac{4 e^{4z}}{(1+2 e^{2z})^3} \,.
\end{eqnarray}
If we take the $\mathsf{R^\ast} \to \infty, \omega \to 0$ limits, the integral on the right hand side is convergent. In order for the left hand side (i.e. the term in  square brackets) to be finite, we must have
\begin{equation} \label{eq:qb_rad}
R\omega^2=1+\sum^{\infty}_{n=0} c_{2n} \omega^{2n+2} \,,
\end{equation}
which confirms the preliminary expectation given by formula \eqref{eq:oom_SBS_R}, i.e. $R^\ast \propto \omega^{-2}, \omega \to 0$. By expanding the denominator for $z \ll \mathsf{R^\ast}$ and ignoring exponentially suppressed terms, we find
\begin{equation}
c_0=\frac{1}{4} (2 \log 2-1)\,, \, c_2=\frac{1}{48} \left(4 \pi ^2-27\right) \label{eq:qb_coeffs}
\end{equation}
for the leading order behaviour.

In order to connect $R^\ast$ to the mass-based radius $R \equiv R_{99}$, we must  numerically  solve $4 \pi \int^{R}_0 dr \, r^2 \rho(r)  = 0.99 M  $
for $R$. Here, we construct an analytic approximate solution by considering the density support of the Q-ball. By construction of the profile,  $\mathsf{R} - \mathsf{R^\ast} \equiv \lambda >0$ and we will  assume that  $\lambda \equiv \mathsf{R} - \mathsf{R^\ast} $ corresponds to the tail of the boundary zone\footnote{The most compact regime corresponds to $\omega \to 0$ when $\lambda \to 0+$. All other, physically reasonable, profiles would present a smoother decay of the scalar i.e. fatter tails. Thus, they would reach the inflection point well before most of $99\%$ of the energy density has been accumulated. We have checked this explicitly, for all of the models considered in this work, both with and without gravity.}.  The width of the boundary zone can be found by taking the derivative of the field \eqref{eq:qb_surface} and computing the standard deviation of that symmetric function, which we generalize by changing $2z \to \delta z$ in the exponent, having in mind the inclusion of gravity in Sec. \ref{sec:gravitating}. This yields
\begin{equation} \label{eq:qb_lam}
\lambda = 3\frac{1}{2}\sqrt{- \int^{+\infty}_{-\infty} z^2 d\Big( \frac{1}{\sqrt{1+2e^{\delta z}}} \Big)} =\frac{3}{2} \frac{2.66}{\delta} \,.
\end{equation}
The prefactor arises from the fact that $\int^{+3\delta}_{-3\delta}d(1+2e^{\delta z})^{-1/2}  \approx 0.99 \int^{+\infty}_{-\infty}d(1+2e^{\delta z})^{-1/2}$. For Q-balls (in Minkowski space) $\delta=2$ and
\begin{equation} \label{eq:qb_rad99}
\mathsf{R} \approx \mathsf{R^\ast} + \lambda  \,, \lambda=2.66 .
\end{equation}

\begin{figure}
\centering
\includegraphics[width=0.42\textwidth]{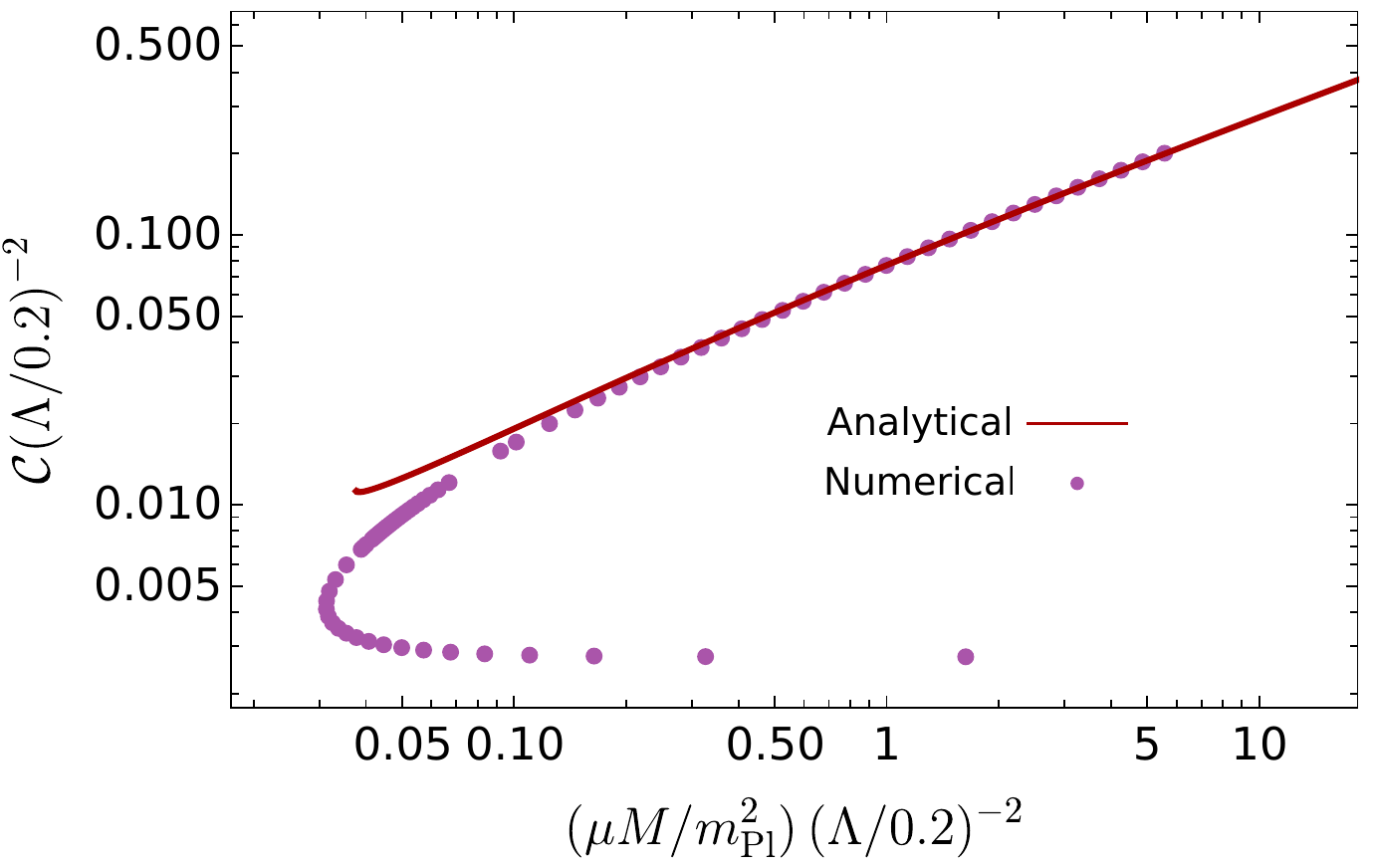}
\caption{Mass-compactness relations for Q-balls. Dots correspond to numerical configurations, while the solid line is found from the analytic expressions \eqref{eq:qb_rad99},  \eqref{eq:qb_M}.}
\label{fig:qb_MR_MC}
\end{figure}

From the solution, other macroscopic parameters can also be calculated, such as the energy (mass) $E=M$ and the Noether charge $Q$ \cite{Heeck:2020bau} :
\begin{eqnarray}
\Bar{Q} &\equiv& \frac{\mu^2 Q}{m^2_{\rm Pl}} \nonumber \\
&\approx&  \frac{\Lambda^2}{6} (\mathsf{R^\ast})^3  w  \Big[1-\frac{3 \ln{2}}{2\mathsf{R^\ast}} + \mathcal{O}{\big( (\mathsf{R^\ast})^{-2} \big)} \Big]\,,   \label{eq:qb_Q} \\
\Bar{M}  &\equiv& \frac{\mu M}{m^2_{\rm Pl}} \nonumber \\
&\approx& w \Bar{Q} +  \frac{\Lambda^2 }{24} (\mathsf{R^\ast})^2 \Big[1+\frac{1- \ln{2}}{\mathsf{R^\ast}} + \mathcal{O}{\big( (\mathsf{R^\ast})^{-2} \big)} \Big]  ,\label{eq:qb_M}
\end{eqnarray}
where we have introduced
\begin{equation}
    \Lambda=\frac{\sigma_0}{M_{\rm Pl}} \,,
\end{equation}
anticipating the connection with the curved space-time case in the next section. Note that the previous relations imply the following scaling of the compactness in the thin-wall regime
\begin{equation} \label{eq:qb_C}
\mathcal{C} \propto \Lambda^2  \frac{1}{\omega^2}  \,.  
\end{equation}
In Fig. \ref{fig:qb_MR_MC} we show the $\mathcal{C} - M$ curve from both the numerical and analytic calculations. From the plots, we also see that the analytic approximation for $R$ agrees well with the numerically calculated value. The strong gravity regime is expected to be attained for values of $\mu \,, M \,, \sigma_0$ such that $\mathcal{C} \geqsim 0.1$:
\begin{eqnarray} \label{eq:qb_M_mu_}
\Big(\frac{\mathcal{{C}}}{0.1} \Big)  &&   \,  \\
&& \approx 1.6 \times \Big(\frac{\mu}{10^{-8} \rm{eV}} \Big)^{1/2} \Big(\frac{M}{1 M_{\odot}} \Big)^{1/2} \Big(\frac{\sigma_0}{10^{17} \rm{GeV}} \Big)^{-1} \nonumber \\
&& \approx 1.6 \times \Big(\frac{\mu}{10^{-14} \rm{eV}} \Big)^{1/2} \Big(\frac{M}{10^6 M_{\odot}} \Big)^{1/2} \Big(\frac{\sigma_0}{10^{17} \rm{GeV}} \Big)^{-1} \,. \nonumber
\end{eqnarray}

The threshold frequency between the stable and the unstable branch can be found from Eqs. \eqref{eq:qb_stab},  \eqref{eq:qb_Q} and   \eqref{eq:qb_M}, and is given by \cite{Heeck:2020bau}
\begin{equation} \label{eq:qb_w_max}
w <  w_{\rm{Qb-s}} \approx  0.82 \,.
\end{equation}

\section{Soliton boson stars}
\label{sec:gravitating}

Turning  gravity on introduces a new scale $\Lambda$  in the problem, with both perturbative and non-perturbative effects. In the low-compactness limit, SBSs effectively ``see'' a quadratic potential and are thus stabilized by  quantum pressure (Sections \ref{sec:oom} and  \ref{sec:par_dilute}). In the high-compactness limit,~Eq. \eqref{eq:qb_C} shows that for any given $\Lambda$ there exists a sufficiently low $\omega$ such that at some point the Schwarzschild compactness will be reached, and the scalar will collapse to a BH, simply because of the hoop conjecture~\cite{Thorne:1972ji, Cardoso:2019rvt}. Thus, from the one stable and the one unstable branch in the flat space-time limit, we expect two stable and two unstable branches for \textit{any} value of $\Lambda \ll 1$. This has
already been established numerically and using  catastrophe theory arguments~\cite{Tamaki:2011zza, Kleihaus:2011sx}. In the following,  we will confirm previous results and supplement them with a physical interpretation and also an analytic model, focusing on the compact stable branch in the perturbative regime $\Lambda \ll 1$.

\subsection{Scaling arguments} \label{sec:oom}

The structure of  macroscopic objects is determined by the physics that stabilizes them. In this section, partially inspired by \cite{Lee:1987rr,Burrows:2014fla,Freivogel:2019mtr}, we will provide rough scaling arguments to demonstrate what kind of  configuration properties are to be expected.

A polytropic equation of state  $P \sim \kappa \rho_{\rm m}^\gamma \,$, $[\kappa]=[M]^{-n}$ provides pressure support counteracting the attractive gravitational force. In  equilibrium, one has
\beq \label{eq:MR_pres}
M^{2-\gamma} \sim \kappa m^2_{\rm Pl}R^{4-3\gamma} \,.
\eeq
If these configurations can reach the Schwarzshild compactness scale $R_{\rm Sch} \sim M/m^2_{\rm Pl}$, where  strong-gravity effects are important,  the
maximum  Chandrasekhar mass is
 given by
\beq
M_{\rm Ch}^{2\gamma-2} \sim \kappa m^{6\gamma-6}_{\rm Pl} \,.
\eeq

SIBSs are pressure-supported by the repulsive interaction $P \sim \lambda |\Phi|^4$ for sufficiently large  values of $\lambda>0$. As the matter density is given by $\rho_{\rm m} \sim \mu^2 |\Phi|^2$ (with $\mu$  the scalar mass), the equation of state is $P \sim \frac{\lambda}{\mu^4} \rho_{\rm m}^2  $.  SIBSs thus have the same mass-radius scaling as fermionic pressure-supported objects [scaling \eqref{eq:MR_pres}], and the  maximum mass is given by the Chandrasekhar limit
\beq
M_{\rm Ch} \sim \sqrt{\lambda} \frac{m^3_{\rm Pl}}{\mu^2}  \,.
\eeq
Pressure-supported nature of the SIBS is also the reason why the maximal attainable compactness for this model $\mathcal{C}_{\rm max}[\rm{SIBS}] \approx 0.16$ \cite{AmaroSeoane:2010qx} is very close to the neutron star value $\mathcal{C}_{\rm max}[\rm{NS}] \approx 0.19$ \cite{Chen:2015zpa}.

The stability of MBSs stems from the interplay of the kinetic term in the Lagrangian (``quantum pressure'') and gravity. Microscopically, quantum pressure originates from Heisenberg's uncertainty principle. Let $R$ be the characteristic size of a MBS and $v_\text{vir}\sim \sqrt{M/(m^2_{\rm Pl}R )}$ its virialized velocity. From the uncertainty principle, one has
\beq
\mu  M \sim \frac{m^2_{\rm Pl}}{\mu R}.
\eeq
The maximum mass, when $R \sim R_{\rm Sch}$, corresponds to  the Kaup limit \cite{Kaup:1968zz}
\beq
M_{\rm Kp} \sim \frac{m^2_{\rm Pl}}{\mu}.
\eeq
Note that the difference between the Kaup and Chandrasekhar scaling originates from the fact that the quantum pressure is  not a polytropic radial pressure, but rather an anisotropic non-local stress \cite{Chavanis:2011cz}.

In contrast to MBSs and SIBSs, SBSs are stabilized already in  flat space-time, as discussed in Sec. \ref{sec:qb}. We can estimate the maximum mass  by equating expression \eqref{eq:oom_SBS_R} with the Schwartzshild scale,
which gives the Lee limit \cite{Friedberg:1986tq, Lee:1987rr}
\beq
M_{\rm Lee} \sim \frac{m^4_{\rm Pl}}{\sigma^2_0\mu}.
\eeq

\subsection{Structure equations in GR} \label{sec:grav_eq}

The action \eqref{eq:qb_action} can be generalized to curved space-time via a minimal coupling to gravity:
\begin{equation} \label{eq:fss_action}
S=\int d^4 x \sqrt{-g} (M^2_{\rm Pl} R + \mathcal{L}_\Phi) \,.
\end{equation}
We will consider spherically symmetric and static space-times of the form
\beq \label{eq:metric_sph}
ds^2=-e^v dt^2 + e^u dr^2 +r^2 d\Omega^2 \,.
\eeq
The radial profile of the metric coefficients follows from the Einstein field equations, which  together, with the Klein-Gordon equation for the scalar, describe the full structure of the object:
\begin{eqnarray}
\frac{1}{r^2}\l(r\,e^{-u}\r)' -\frac{1}{r^2}&=& -\frac{1}{M^2_{\rm Pl}} \rho\,,\label{eq:u_struc}\\
e^{-u}\l(\frac{v'}{r}+\frac{1}{r^2}\r)-\frac{1}{r^2}&=&\frac{1}{M^2_{\rm Pl}} P_{\text{rad}}\,,\label{eq:v_struc}\\
\phi''+\l(\frac{2}{r}+\frac{v'-u'}{2}\r)\phi'&=&e^u\l({\frac{dV}{d|\Phi|^2}}-\omega^2e^{-v}\r)\phi,\label{eq:SBS_KG}
\end{eqnarray}
with the density $\rho$, the radial pressure $P_{\rm rad}$ and the tangential pressure $P_{\rm tan}$ defined by
\begin{eqnarray}
\rho=e^{-v} \omega ^2  \phi^2 + e^{-u}
(\phi') ^2 + V   \,,  \label{eq:sbs_rho} \\
P_{\rm rad} = e^{-v} \omega ^2  \phi^2 + e^{-u} (\phi') ^2-V   \,, \label{eq:sbs_prad} \\
P_{\rm tan} =e^{-v}  \omega ^2  \phi^2 - e^{-u} (\phi') ^2 - V   \,.  \label{eq:sbs_ptan}
\end{eqnarray}

We will now focus on the simplest potential  \eqref{eq:V6qball} and adopt the rescaling  
\begin{equation}
\mathsf{r} = \mu r \,, \Bar{m}(\mathsf{r})=\frac{\mu m(r)}{m^2_{\rm Pl}}  \,, w=\frac{\omega}{\mu}  \,, \varphi=\phi/(\sigma_0/\sqrt{2})
\end{equation}
[c.f. ~relations \eqref{eq:qb_scale}].
With this parameterization, the structure equations become dimensionless. We further assume that $' \equiv d/d\mathsf{r}$ unless we state otherwise.

Note that a shift $v \to \tilde{v}=v - v(0)$ corresponds to
a rescaling of the time coordinate and thus to
a redefinition of the scalar frequency $\tilde{\omega}=\exp{(-v(0)/2)}\omega$.  Fixing $\varphi(0) \equiv \varphi_c$ and $\varphi(\infty)=0$,
let us use this gauge freedom to also set $\tilde{v}(0)=0$. This
specifies\footnote{We are here focusing on the ground state (nodeless) solutions. It has been previously established that the excited solutions in various boson star models decay to the ground state \cite{Balakrishna:1997ej,Collodel:2017biu}. However, there are recent indications that sufficiently strong self-interactions can stabilise excited states over long time scales  at least for SIBSs \cite{Sanchis-Gual:2021phr}.} a boundary value problem, with  eigenvalue $\tilde{\omega}$, which we determine with a shooting method. Once that the configuration is calculated in this gauge, we can rescale the time coordinate at will, so that $v(\infty) \equiv v_\infty =0$. More details on the numerical method are given in the App. \ref{AppEKG}.

The Arnowitt-Deser-Misner (ADM) mass \cite{carrollbook} can be extracted from the asymptotic behaviour of the metric functions:
\begin{equation} \label{eq:SBS_Sch_mass}
\exp{[-u(\mathsf{r})]}= 1-\frac{2\Bar{M}}{\mathsf{r}} + \mathcal{O} \Big( \frac{1}{\mathsf{r}^2} \Big) \,.
\end{equation}
From Eq. \eqref{eq:v_struc}, one can also obtain the integral representation
\begin{equation} \label{eq:SBS_M_int}
\Bar{M}=4 \pi \int d \mathsf{r} \, \mathsf{r}^2 \frac{\rho(\mathsf{r}')}{\mu^2 \sigma_0^2} \,.
\end{equation}
Finally, the Noether charge is given by
\begin{eqnarray}
Q &\equiv& - \int dV j^t = 4 \pi \omega \int dr d\theta \, \sqrt{-g} e^{-v}  \phi^2 \,. \label{eq:SBS_charge_GR}
\end{eqnarray}

\subsection{Representative configurations}

In the rest of this Section, we will set  $\Lambda=0.186$ in order to illustrate our results. In the gauge $\tilde{v}(0)=0$, the fictitious particle initially does not feel the impact of gravity,  and consequently, in the thin-wall regime, we should expect that the central field is close to $\varphi_+$.

\begin{figure*}[th]
\includegraphics[width=0.95\textwidth]{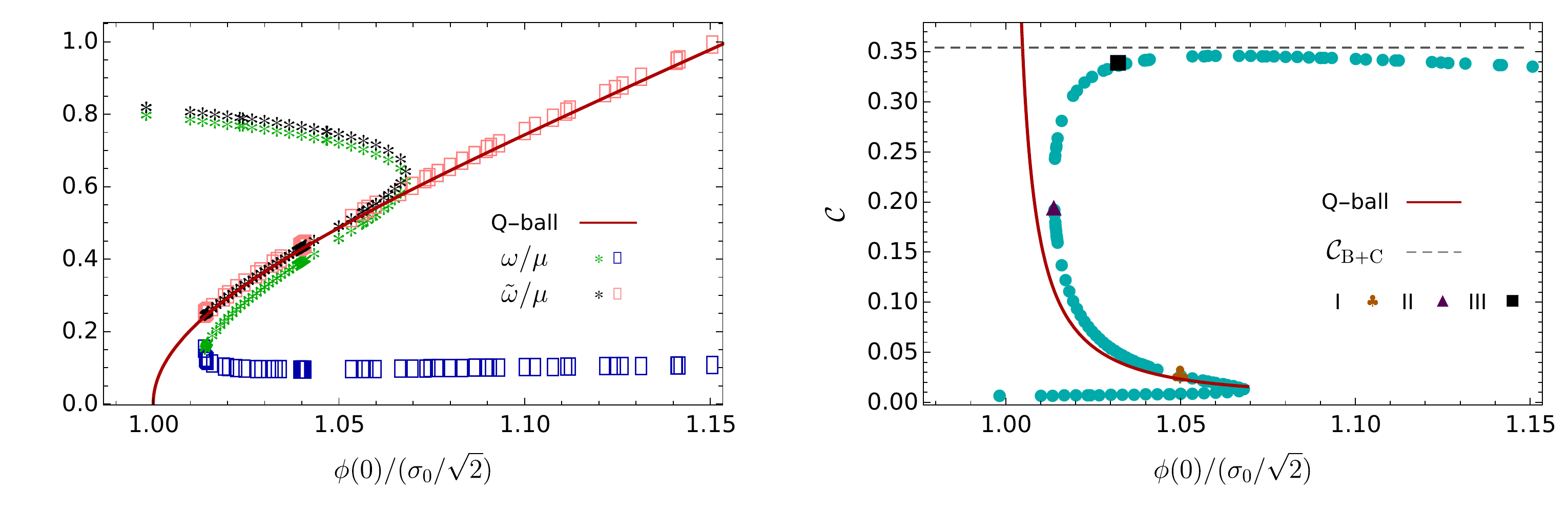}
\caption{(Left) Frequency vs central field $\varphi_c \equiv \phi(0)/(\sigma_0/\sqrt{2})$ for $\Lambda=0.186$, in two gauges $\omega \,, \Tilde{\omega}$ compared to the Q-ball prediction appropriate in the $\Tilde{v}(0)=0$ gauge (dark red line). Note that two almost identical segments of the thin-wall regime in the $\Tilde{v}(0)=0$ gauge, denoted by the pink rectangles and the black stars, correspond to separate segments in the $v(\infty)=0$ gauge, denoted by the blue rectangles and the green  stars, respectively.   (Right) Compactness vs central field  for $\Lambda=0.186$ (cyan circles) and the Q-ball result (dark red line). The three representative configurations (denoted by  the  orange club for \rm{I}, purple triangle for \rm{II} and black square  for \rm{III}) are indicated.}
    \label{fig:sbs_representative_lam}
\end{figure*}

Like in the flat space-time case, in curved space-time numerical results produce two types of configurations - thin-wall ones for small $w$, and  thick-wall ones for larger $w$. Unlike in the  Q-ball case, the thin-wall regime of SBSs has two sub-classes, one perturbatively close to the flat space-time case, and a non-perturbative one close to the maximum compactness configuration. In Fig. \ref{fig:sbs_representative_lam} we show the $w  - \varphi_c$, $\tilde{w}  - \varphi_c$ and $\mathcal{C} - \varphi_c$ curves for our choice of $\Lambda$.

Various authors have used catastrophe theory arguments to assess the (local) stability of Q-balls and boson stars \cite{Sakai:2007ft, Tamaki:2010zz, Tamaki:2011zza, Kleihaus:2011sx}. We will not review these arguments here, but let us mention that they show that  stability can be checked by analyzing the position of the turning points, for fixed $\Lambda$, in the $\Bar{M} - \varphi_c$ curve, up to a final (unstable) ``spiral'' that occurs for $\varphi_c \geqsim 1$ in the gravitating case. Thus, one can start from the stable MBS limit and track the turning points as  $\varphi_c$ increases. Already from Fig. \ref{fig:sbs_representative_lam}, we see two turning points that mark the boundaries of a middle stable branch, with two unstable branches on the left and on the right of it (a low compactness stable branch is not shown on this plot, see Fig. \ref{fig:v4_fi_M}). Note however that the turning point in the $\Bar{M} - \varphi_c$ parameter space does not map exactly to the same point in the $\mathcal{C} - \varphi_c$ parameter space. We have established that the most massive configurations, which separate the compact stable branch from the unstable one, have compactnesses slightly bellow the maximum one. For example, for our representative case, the maximum mass configuration has $\mathcal{C}=0.336$, while the maximally compact one has $\mathcal{C}=0.346$.

We will now focus on the three representative configurations for each of the three aforementioned (sub-)classes: the thick-wall regime (\rm{I}); the thin-wall regime perturbatively close to the Q-ball limit (\rm{II}); and the thin-wall regime close to the non-perturbative $\mathcal{C}_{\rm max}$ cut-off  (\rm{III}) for our $\Lambda=0.186$. Configuration III is also the maximally massive one for our $\Lambda$. The parameters of these configurations are given in Table \ref{tab:1}. The field and density profiles are shown in Figs. \ref{fig:sbs_cs}  and \ref{fig:sbs_asymp_bound}, while the metric coefficients are displayed in Fig.~\ref{fig:sbs_u_v_II}.

\begin{table*} 
 \begin{tabular}{||c c c c c c c c||}
 \hline
 & $\phi(0)/(\sigma_0/\sqrt{2})$ & $\Tilde{\omega}/\mu$  & $\omega/\mu$ & $\tilde{v}(\infty)$ & $\mu R$ & $\mu M/m^2_{\rm Pl}$  & $\mathcal{C}$     \\[0.5ex]
\hline\hline
\hline
I & 1.05 & 0.488 & 0.454 & 0.147 & 7.243 & 0.188 & 0.026 \\
\hline
II & 1.014 & 0.241 & 0.146 & 1.009 & 29.821 & 5.719 & 0.192   \\
\hline
III & 1.032 & 0.380 & 0.088 &  2.924 &  41.182 &  13.84 & 0.336 \\
\hline \end{tabular} \caption{Parameters of the representative configurations of SBSs: thick-wall regime (\rm{I}); thin-wall regime perturbatively close to the Q-ball limit (\rm{II});  strong gravity,  thin-wall non-perturbative branch (\rm{III}). Configuration III has also the maximal value of mass for $\Lambda=0.186$ and thus is at the border between the compact stable and the unstable branch.} \label{tab:1}
\end{table*}

\subsection{Origin of maximum compactness} \label{sec:sbs_maxC}

The maximum compactness of an object is usually discussed in the context of  Buchdahl's theorem \cite{Buchdahl:1959zz}. The latter states, in its most general form, that GR self-gravitating objects with   energy density monotonically decreasing outwards $\rho' \leq 0$ and positive (or vanishing) pressure anisotropy (i.e.  radial pressure  larger than or equal to the tangential one) $P_{\rm rad} \geq P_{\rm tan}$  must have $C \leq 4/9$~\cite{Buchdahl:1959zz,Urbano:2018nrs}. In the case of fluid stars, isotropic ``normal matter'' (i.e. one satisfying the weak energy condition $ \rho \geq 0 \, \land \, \rho + P \geq 0 $ and the micro-stability condition $P \geq 0 \, \land \, dP/d\rho \geq 0$) is consistent with these assumptions. However,  Buchdahl's limit is not a robust concept, as it can be easily evaded by violating the theorem's assumptions,
producing even more compact objects. A discussion of the parameterized bounds on $\mathcal{C}$ can be found in \cite{Guven:1999wm},  and various toy models that exceed the Buchdahl bound have been constructed in the literature \cite{Cardoso:2019rvt}.

Buchdahl's theorem can be additionally strengthened by requiring that the equation of state be consistent with causality. In \cite{Urbano:2018nrs} (see also \cite{Alho:2021sli}), it has been shown that a useful toy model for understanding the maximally compact and causal configurations is given by objects described by a linear equation of state (LinEoS)
\beq
\rho=\rho_c + \frac{P}{c^2_s} \,,
\eeq
where $c^2_s \equiv \partial P/\partial \rho>0$ (the speed of sound) and $\rho_c$ are constant.  Hence, allowing for $c^2_s>1$ accounts for violations of causality. The limit $c^2_s \to \infty$ corresponds to  constant density stars, whose maximum compactness $\mathcal{C}_{\rm B}=4/9$ follows from Buchdahl's theorem. For a given $c_s^2$, the LinEoS describes the stiffest possible matter, and consequently it yields the most compact configurations. Maximally compact and causal configurations consistent with the assumptions of  Buchdahl's theorem cannot surpass  $\mathcal{C}_{\rm{B+C}}=0.354$ when $c^2_s=1$. The dependence of the compactness of LinEoS configurations on $c^2_s$ is approximately given (to within a $3.6\%$ error)  by the fitting formula \cite{Urbano:2018nrs}
\begin{equation} \label{eq:LinEos_C_w}
\frac{4}{9\mathcal{C}_{\rm{LinEoS}}} -1 \approx \frac{0.77+0.51 c^2_s}{c^2_s (4.18 + c^2_s)} \,.
\end{equation}

SBSs in the thin-wall regime are a physical example of objects with a LinEoS\footnote{This is, as far as we know, original insight. A comparison between  SBSs and  constant density stars $(c^2_s \to \infty)$ was discussed in \cite{Macedo:2013jja}.}, because  in the bulk of the star $\varphi \approx 1$  and hence $\varphi' \approx V \approx 0$, making in turn  $  P_{\rm rad} \approx \rho$.   This argument implies that the maximal compactness of SBS is $\mathcal{C}_{\rm max} \leqsim \mathcal{C}_{\rm B+C}$, which is consistent with our numerical results and the values reported in the previous work \cite{Friedberg:1986tq,Kesden:2004qx,Cardoso:2016oxy,Visinelli:2021uve}. Now we address two apparent loopholes in the previous argument. First,
SBSs do {\it not} have monotonic energy density profiles in their surface regions (Fig.     \ref{fig:sbs_asymp_bound}, right),
which violates the assumptions of Buchdahl's theorem~\cite{Urbano:2018nrs}.
The region where this violation occurs, however, is parametrically smaller than the size of the SBS bulk in the thin-wall regime, and thus it does not affect  $\mathcal{C}_{\rm{B+C}}$ significantly. Secondly, although SBSs have anisotropic pressure,
the radial pressure is larger than the tangential one [Eqns. \eqref{eq:sbs_prad}, \eqref{eq:sbs_ptan}],
which does not allow for violating the Buchdahl compactness bound (unlike the opposite case in which the tangential pressure is larger than the radial one \cite{Raposo:2018rjn,Alho:2021sli}).

In more detail, we can estimate how well the LinEoS describes the matter inside the SBS in the thick wall regime. Assuming the Q-ball results 
(c.f. Section \ref{sec:sbs_an} for a justification) we will ignore exponentially suppressed scalar derivatives [Eq. \eqref{eq:qb_c<}] and approximate Eqns. \eqref{eq:sbs_rho}, \eqref{eq:sbs_prad} to obtain
\begin{eqnarray}
\frac{\rho}{\frac{\mu^2 \sigma^2_0}{2}}  \approx w_+^2 \varphi^2_c + V(\varphi_c) \,, \frac{P_{\rm rad}}{\frac{\mu^2 \sigma^2_0}{2}} \approx w_+^2 \varphi^2_c - V(\varphi_c) \nonumber \,,
\end{eqnarray}
from where it follows that
\begin{eqnarray}
(c^2_s)_a \approx \frac{\varphi_c^2 \left(3 \varphi_c^2-2\right)}{6 \varphi_c^4-6 \varphi_c^2+1} \,, \label{eq:linEoS_cs_qb}
\end{eqnarray}
with $w_+$ being the inverse of Eq. \eqref{eq:qb_central} and $\varphi_+ \approx \varphi_c$. For $\varphi_c = 1.032$ (configuration III) from Eq. \eqref{eq:linEoS_cs_qb} we find $(c_s)_a = 0.95$. Using this value in Eq. \eqref{eq:LinEos_C_w} we get $\mathcal{C}_{\rm LinEoS} = 0.350$, close to $\mathcal{C}_{\rm III} = 0.336$.  Taking the limit $\varphi_c \to 1$, Eq. \eqref{eq:linEoS_cs_qb}  implies $(c^2_s)_a \to 1$ and hence $\mathcal{C}_{\rm max} \to \mathcal{C}_{\rm B+C}$. Note however that the exact limit $\varphi_c = 1$ in the thin-wall regime is attainable only in the Minkowski space-time ($\Lambda=0$). This is a singular limit as the absence of gravity implies $\mathcal{C} \to \infty$ when $
\varphi \to 1$, as elaborated in Section \ref{sec:qb_simp}.

\begin{figure}
\centering
\includegraphics[width=0.42\textwidth]{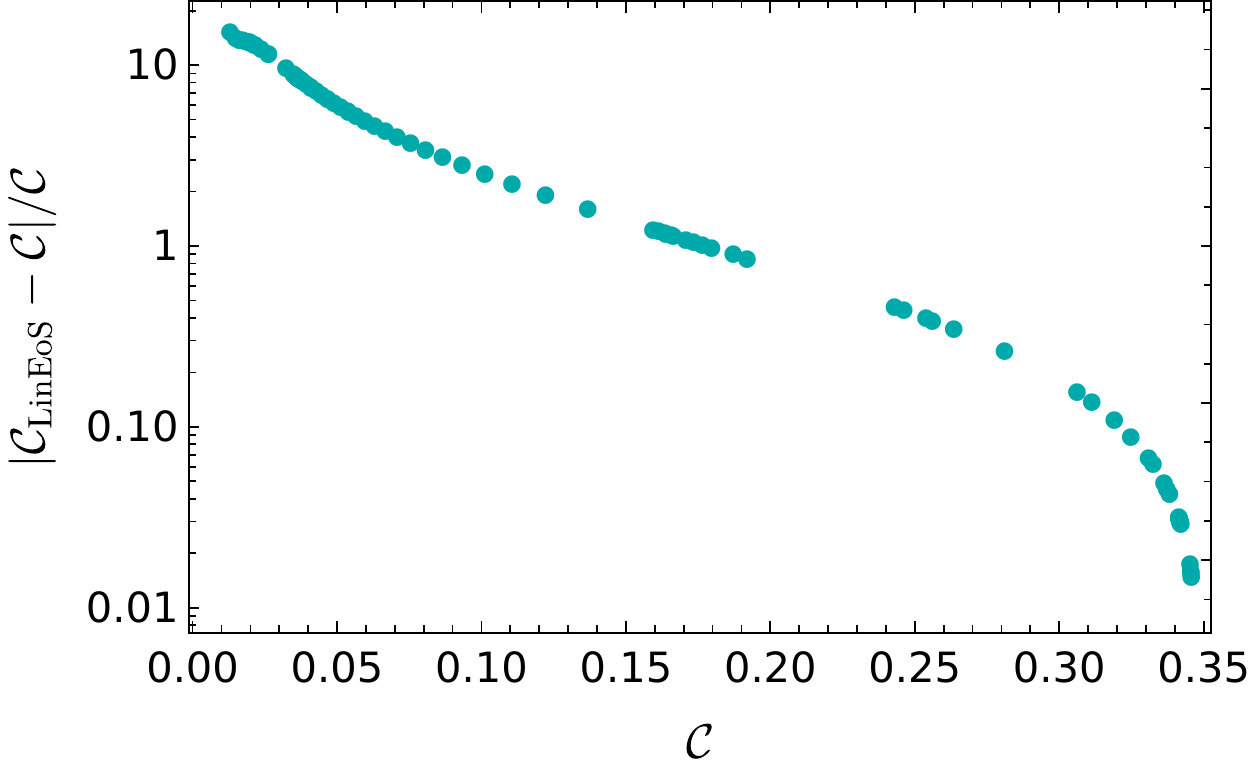}
\caption{Relative difference between the numerically determined compactness $\mathcal{C}$ and the prediction from the LinEoS $\mathcal{C}_{\rm LinEoS}$ \eqref{eq:LinEos_C_w} [using the numerically found average speed of sound  \eqref{eq:LinEos_C_w_av}].}
\label{fig:sbs_C_comp}
\end{figure}

\begin{figure}
\centering
\includegraphics[width=0.4\textwidth]{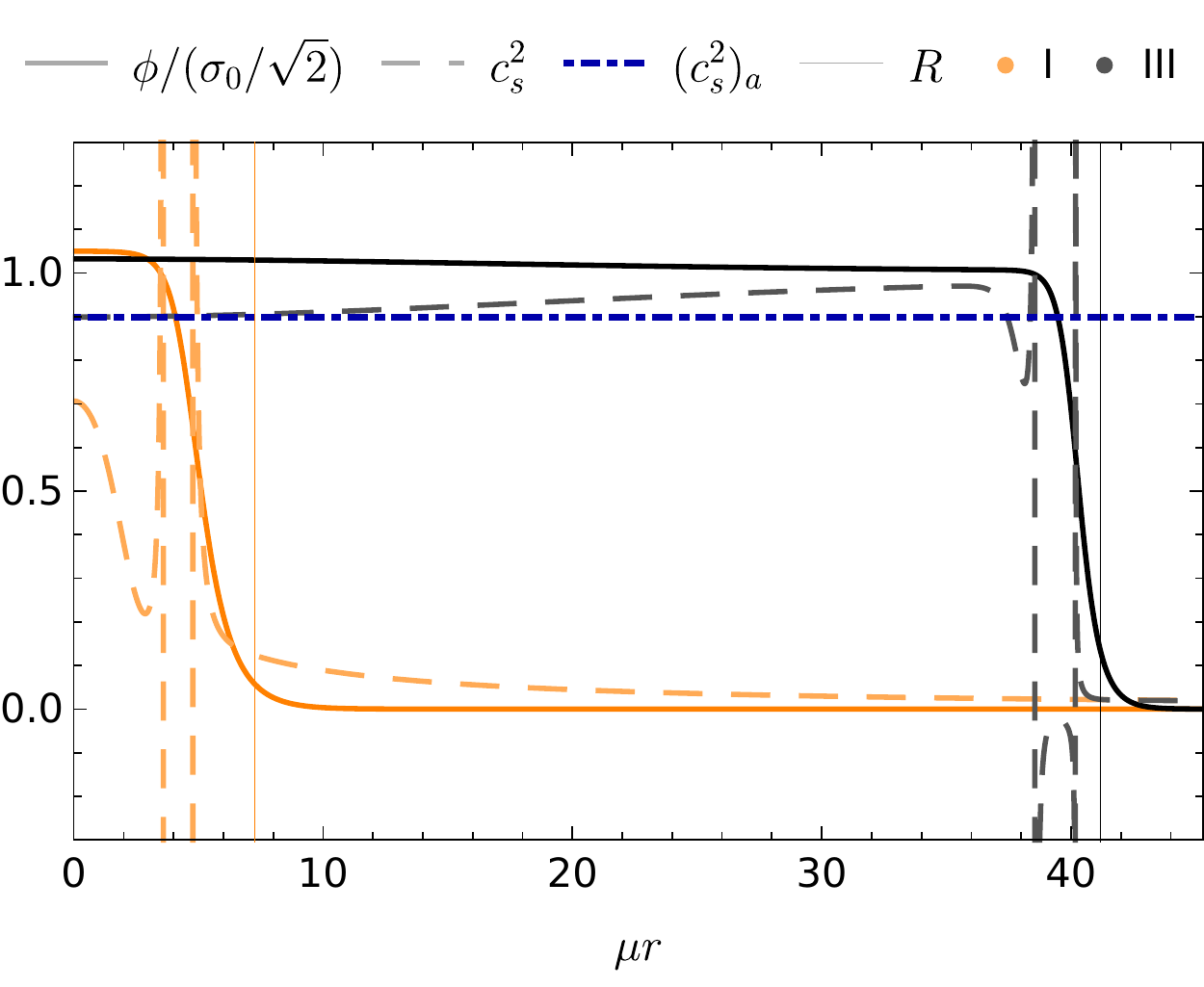}
\caption{Scalar field $\varphi \equiv \phi/(\sigma_0/\sqrt{2})$  and  the speed of sound $c^2_s$ radial profile for the two representative configurations: \rm{I} (orange) and \rm{III} (black). The blue dot-dashed line represents the  estimate of the speed of sound (for III) $(c^2_s)_a$ from Eq. \eqref{eq:linEoS_cs_qb}.}
\label{fig:sbs_cs}
\end{figure}

In order to scrutinize previous analysis further, we have calculated (for our representative $\Lambda= 0.186$) numerically the average
\begin{equation} \label{eq:LinEos_C_w_av}
\langle c^2_s \rangle =\frac{1}{R_<} \int^{R_<}_0 dr \frac{\partial P_{\rm rad}}{\partial \rho}
\end{equation}
over several configurations [with $R_<$  the boundary of the bulk, c.f. expression \eqref{eq:sbs_bulk_size}], and we have
 compared the compactness predicted by Eq. \eqref{eq:LinEos_C_w} with the actual one. The results are shown in Fig.  \ref{fig:sbs_C_comp} , and clearly indicate
 good agreement
 in the thin wall regime (i.e. at high compactnesses). We
  have also presented results for the
 position dependence of $c^2_s$ in Fig. \ref{fig:sbs_cs} ,  for two specific configurations  (\rm{I} and \rm{III}). Note that negative and seemingly non-causal values of $c^2_s$ appear in the boundary zone,  signaling a breakdown of the hydrodynamic description. This breakdown also occurs in  flat space-time~\cite{Nugaev:2019vru}.

Finally, in the thick-wall regime, commensurability between the bulk and the boundary does not allow for an effective LinEoS description (Fig. \ref{fig:sbs_cs}). As a result, in this regime  $\mathcal{C}\ll\mathcal{C}_{\rm{B+C}}$.

\subsection{Analytic construction} \label{sec:sbs_an}

As already mentioned, the structure of SBSs can also be interpreted  in light of the dynamics of a fictitious Newtonian particle in a time dependent potential, with $r$ playing the role of time\footnote{This interpretation was mentioned in \cite{Tamaki:2011bx} at the qualitative level, but without exploring its implementation and  consequences. A similar approach  was also applied to (Minkowski) Q-balls with gauged $U(1)$ symmetry while this work was well into preparation \cite{Heeck:2021zvk}. }, as can be seen from the Klein-Gordon equation
\begin{eqnarray}
&& \varphi''+\Big( \frac{ 2}{\mathsf{r}}- \frac{W'}{W}  \Big) \varphi' = \Big[\mathsf{m} ^2(1-4\varphi^2+3\varphi^4)-W^2 \Big] \varphi \,,  \nonumber\\
&& \mu W = \omega e^{(u-v)/2}  \,, \mu \mathsf{m} = \mu e^{u/2} \,,\label{eq:KG_time}
\end{eqnarray}
Equations  \eqref{eq:u_struc} and \eqref{eq:v_struc} can be formulated as dynamical equations for ``time-dependent'' frequency $\mu W $ and scalar mass $\mu \mathsf{m}$. Their dynamics depends on $\varphi$ through  $\rho$ and $P_{\rm rad}$. However, in certain regimes we can approximately evaluate  $W,\mathsf{m}   $ without knowing the full behaviour of $\varphi$.

If we work in the gauge where $\tilde{v}(0)=0$, the Q-ball does not ``feel'' the gravitational field initially (i.e. at the centre).  Like in the flat space-time case, for a given $\varphi_c$ we need  to find $W(0)=w$  such that the ``SBS-particle'' rolls over the time-dependent potential and reaches $\varphi=0$ in an infinite time ($\mathsf{r}\to\infty$) with $\mathcal{E}=0$.  Note that  $W$ is a gauge invariant object (i.e. it is left unaffected by a change of the time coordinate).

The particle analogy sheds light, already at the qualitative level, on some of the properties of SBSs: as Q-balls have a unique thin-wall regime, for SBSs this regime will be  described  by a curve in the $\tilde{w}  - \varphi_c$ space, very close to the Q-ball result. However, for some SBS configurations, the effective particle will exhibit the thin-wall regime perturbatively close to the Q-ball one (like for configuration II), while  others will be in the non-perturbative part of the parameter space (like in the case of  configuration III). Thus, the thin wall regime in the $\tilde{w}  - \varphi_c$ representation consists of two almost degenerate curves (denoted by black stars and pink rectangles in Fig. \ref{fig:sbs_representative_lam}). For example, there are configurations where the difference in $\tilde{w}$ can be as low as $\sim 10^{-51}$, with the corresponding difference in $w$ $\sim 10^{-6}$. These curves split into two separate curves in the $w  - \varphi_c$ parameter space, one close to the Q-ball line, while the other is the horizontal asymptote, represented by  green stars and  blue rectangles on Fig. \ref{fig:sbs_representative_lam}, respectively.

In the following subsections, we will give an analytic description of the scalar in the strong-field regime in three zones (analogous to the flat-spacetimes ones introduced in Sec. \ref{sec:qb}), starting from the simplest one.

\subsubsection{Exterior zone}  \label{sec:sbs_asymp}

The simplest regime is the asymptotic one, where the space-time is to a good approximation Schwarzschild, owing to the fast decay of the scalar:
\beq
u_>=-\log{\Big(1-\frac{2\Bar{M}}{\mathsf{r}} \Big)}\,, \label{eq:u_asymp} \\
\tilde{v}_>=\tilde{v}_\infty+\log{\Big(1-\frac{2\Bar{M}}{\mathsf{r}} \Big)}\, \label{eq:v_asymp}.
\eeq
The evolution of the scalar is found by solving the Klein-Gordon equation \eqref{eq:SBS_KG} with these $u, v$,  expanding in powers of $1/r$:
\beq
&& [\mathcal{O}_{\rm L}+\mathcal{O}_{\rm NL}]\varphi(\mathsf{r}) = 0  \,, \\
&& \mathcal{O}_{\rm L} = \frac{d^2}{d\mathsf{r}^2}+\frac{2}{\mathsf{r}}\frac{d}{d\mathsf{r}}- \Big(1-e^{-\tilde{v}_\infty} \tilde{w} ^2 +\frac{2\Bar{M}}{\mathsf{r}} \big( 1-2e^{-\tilde{v}_\infty} \tilde{w}^2 \big)\Big)  \nonumber \,, \\
&& \mathcal{O}_{\rm NL} = \frac{\Bar{M}}{\mathsf{r}} \Big(6   \varphi ^5-8   \varphi^3 \Big) +3 \varphi^5-4   \varphi^3 \nonumber \,.
\eeq
Note that the non-linear terms can be treated as a perturbation. While it appears that there is no simple analytic estimate of these corrections, the leading term gives a good description of the asymptotic behaviour,
because the scalar is suppressed (more than exponentially).

\begin{figure*}[th]
\includegraphics[width=\textwidth]{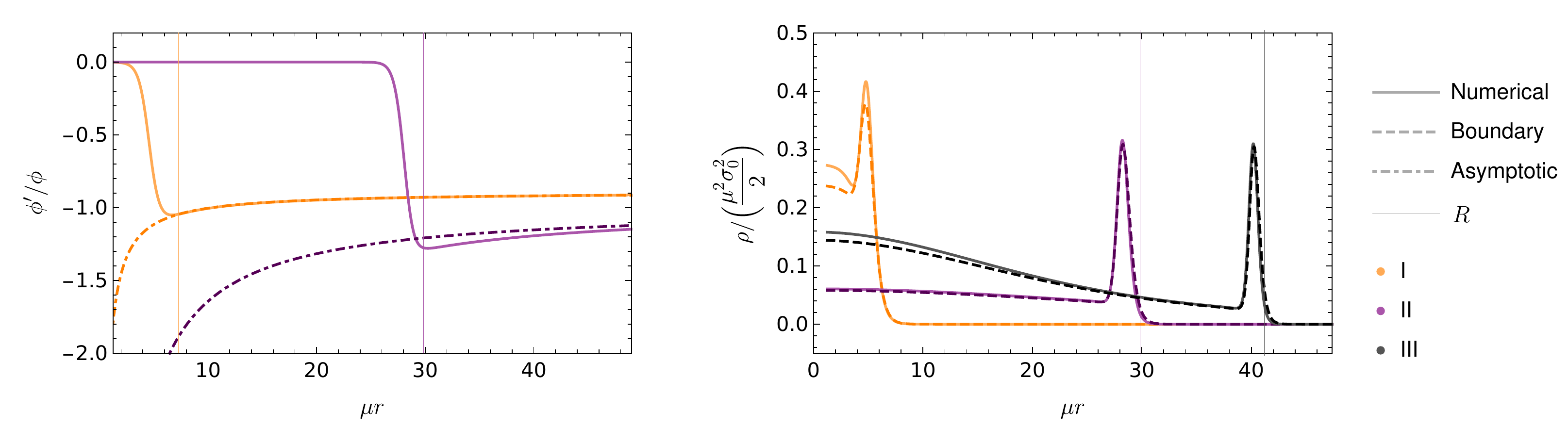}
\caption{(Left) Numerical results vs asymptotic approximation for the ratio between the field radial profile and its derivative $\phi/\phi'$ for configurations \rm{I}, \rm{II}. (Right) Numerical results vs boundary zone analytic approximation for the field energy density radial profile $\rho/(\frac{\mu^2 \sigma^2_0}{2})$ for configurations \rm{I}, \rm{II}, \rm{III}. Numerically determined parameters are used as  input.}
    \label{fig:sbs_asymp_bound}
\end{figure*}

Requiring $\varphi\to0$ as $r\to\infty$, we obtain the solution in terms of hypergeometric functions, or equivalently in terms of the Wittaker function~\cite{Kling:2017mif}
\beq \label{eq:SBS_asymp}
\varphi_{\infty}=\frac{\mathcal{A}}{2 \mathsf{r} \sqrt{1-e^{-\tilde{v}_\infty}\tilde{w}^2 }} \times \nonumber \\ \,W\Big[\Bar{M}\frac{ 2 e^{-\tilde{v}_\infty} \tilde{w} ^2-1}{\sqrt{1-e^{-\tilde{v}_\infty} \tilde{w} ^2}},-\frac{1}{2},2 \sqrt{1-e^{-\tilde{v}_\infty} \tilde{w} ^2} \Big] \,.
\eeq
This asymptotic solution is parameterized by $\tilde{v}_\infty$, $\Bar{M}$, $\tilde{w}$ and the normalization amplitude $\mathcal{A}$. The linear equation $\mathcal{O}_{\rm L}\varphi(\mathsf{r})=0$ is invariant under the rescaling $\varphi \to \mathsf{C} \varphi$, but the matching condition breaks the invariance and selects $\mathsf{C}=\mathcal{A}$. Expanding the Whittaker function in $1/\mathsf{r}$, the leading behaviour gives
\begin{eqnarray} \label{eq:asymp_LO}
\varphi_{\infty} \simeq \frac{\mathcal{A}}{\mathsf{r} ^{1+\beta_>}}e^{-\alpha_> \mathsf{r} }  \,, \\
\alpha_> =\sqrt{ 1-e^{-\tilde{v}_\infty} \tilde{w}^2 } \,, \\
\beta_> = \frac{\Bar{M}}{\alpha_>}(1-2 e^{-\tilde{v}_\infty} \tilde{w}^2) \,.
\end{eqnarray}
Note that for $\Lambda \to 0$, $\Bar{M} \to 0$ and we recover the Q-ball result  \eqref{eq:qb_asymp}. We show a comparison between the numerical and the asymptotic field behaviour in Fig. \ref{fig:sbs_asymp_bound}.

\subsubsection{Interior zone} \label{sec:inteior}

In the gauge $\tilde{v}(0)=0$, the  flat space-time result  \eqref{eq:qb_c<} should be valid sufficiently close to the origin, and the scalar field derivative is suppressed by $\tilde{w}$. Thus, we will  calculate the metric coefficients in the interior perturbatively in $\tilde{w}$, and approximate $\varphi' \approx 0$ and $ V_< \approx  V(\varphi_c) $. This description provides a good approximation for  configurations similar to \rm{II} in the strong-field and the thin-wall regime and close to the Q-balls limit, where $\Tilde{w}$ is small and  controls the size of the star.

\begin{figure*}[th]
\includegraphics[width=\textwidth]{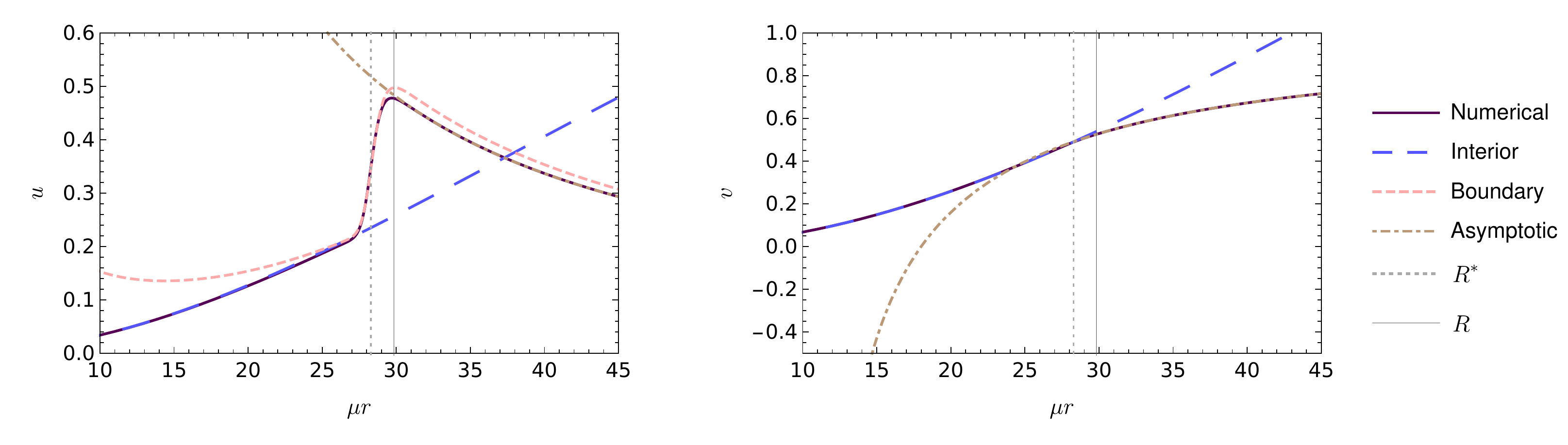}
\caption{(Left) Numerical vs. approximate analytic (interior, boundary, asymptotic) results for the $u$ metric coefficient radial profile for \rm{II}.  (Right) Numerical vs. approximate analytic (interior, asymptotic) results for the $v$ radial profile metric coefficient for \rm{II}. Numerically determined parameters are used as  input.}
    \label{fig:sbs_u_v_II}
\end{figure*}

The perturbative expansion in $\tilde{w}$ must be appropriately resummed (or performed from the start in a suitable form) so that it can be matched with the exterior. Taking
\begin{eqnarray}
\tilde{u}_<=\log[\mathsf{\tilde{u}}_0(r)+\mathsf{\tilde{u}}_2(r) \tilde{w}^2 + \mathsf{\tilde{u}}_4(r) \tilde{w}^4 + \mathcal{O}(\tilde{w}^6)]    \label{eq:u_int} \,, \\
\tilde{v}_<=\log[\mathsf{\tilde{v}}_0(r)+ \mathsf{\tilde{v}}_2(r) \tilde{w}^2 + \mathsf{\tilde{v}}_4(r) \tilde{w}^4 + \mathcal{O}(\tilde{w}^6)]  \label{eq:v_int} \,,
\end{eqnarray}
from Eqns. \eqref{eq:u_struc} and \eqref{eq:v_struc} one gets
\beq
\mathsf{\tilde{u}}_0=1   \, , \,\, \mathsf{\tilde{u}}_2=\frac{1}{6}\Lambda ^2 r^2  \, ,   \mathsf{\tilde{u}}_4= \frac{1}{360} \left(45 \Lambda ^2 r^2-2 \Lambda ^4 r^4\right)\, \\
\mathsf{\tilde{v}}_0=1   \,, \mathsf{\tilde{v}}_2=\frac{1}{3}\Lambda ^2 r^2  \,, \mathsf{\tilde{v}}_4=\frac{1}{120} \left(15 \Lambda ^2 r^2 + 4 \Lambda ^2 r^4 \right)\,.
\eeq
In Fig. \ref{fig:sbs_u_v_II} we show the numerical results and the interior and asymptotic approximations for the metric coefficients.

\subsubsection{Boundary zone} \label{sec:ana_boundary}

In the transition region, let us expand around $(\mathsf{R^\ast})^{-1}$ as in \cite{Heeck:2020bau}. Thus, we neglect the friction $1/R^\ast$ term in the Klein-Gordon equation  \eqref{eq:SBS_KG}. We can estimate the contribution from the time-dependent frequency as
\begin{equation}
\frac{W'}{W} \sim \frac{1}{1-\frac{2\Bar{M}}{\mathsf{R^\ast}}} \frac{2\Bar{M}}{(\mathsf{R^\ast})^2} \sim \frac{2}{\mathsf{R^\ast}(\mathcal{C}^{-1}-1)} \,.
\end{equation}
As $\mathcal{C}_{\rm{max}} \sim 0.35$ (cf. Sec. \ref{sec:sbs_maxC}), we find that this term is of the order of $1/R^\ast$ and can be neglected. Now, neglecting friction and assuming in the first iteration $W \approx W_\ast \equiv W(\mathsf{R^\ast}) \,, \mathsf{m}  \approx \mathsf{m} _\ast \equiv \mathsf{m} (\mathsf{R^\ast})$ (as we are interested only in a tiny strip around of the thin wall), we can use the conservation of  energy and the fact that $\mathcal{E} (\infty)=0$ to reduce the equations of motion to~\cite{Heeck:2020bau}
\begin{equation}
\varphi'= \pm \varphi \sqrt{\Big[\mathsf{m} ^2_\ast (1-\varphi^2)^2-W^2_\ast \Big]}  \,.
\end{equation}
In the thin-wall regime, the expectation $W^2_\ast \ll 1$ leads to
\begin{equation} \label{eq:sbs_surface}
\varphi_{B} = \frac{1}{\sqrt{1+2\exp{[2\mathsf{m} _\ast(\mathsf{r}-\mathsf{R^\ast})]}}}  \,,
\end{equation}
where we have specified the integration constant by requiring $\varphi''(\mathsf{R^\ast})=0$. Note that the second integration constant is determined by the value of the energy and the requirement that the function is monotonously decreasing. Furthermore, \cite{Heeck:2020bau} includes an {\it ad hoc} prefactor $\varphi_+$ in $\varphi_{B}$ for Q-balls [formula \eqref{eq:qb_surface}], because it slightly improves the analytic description in the thick-wall limit. (Note that in the thin-wall regime $\varphi_+ \approx 1$ in any case.) We have not included this prefactor in the SBS context, as it tends to worsen the model.

From the analysis around  Eq. \eqref{eq:qb_lam}, we find
the estimate of the width of the density support in the boundary zone to be
\begin{eqnarray}
\lambda = \frac{3}{4} \frac{2.66}{\mathsf{m}_\ast} \,, \\
\mathsf{R_{>/<}} = \mathsf{R^\ast} \pm \lambda \label{eq:sbs_bulk_size}
\end{eqnarray}
In the flat space-time limit $\mathsf{m}_\ast=1$ and we recover Eq. \eqref{eq:qb_surface}.

A plot of the field density is given in Fig.  \ref{fig:sbs_asymp_bound}, where the metric coefficients are taken from the interior zone perturbative series \eqref{eq:u_int},  \eqref{eq:v_int}. Note that the
boundary-zone solution works (somewhat surprisingly) even far from its region of a priori validity, down to the transition region (like in the flat space-time case). The worst agreement occurs for the configuration \rm{I}, as expected since this configuration is in the thick-wall regime.

Having a preliminary understanding of the field behaviour in the boundary zone, as well as of the radial profile of the metric coefficients in the internal and  external zones, we can understand the junction conditions for the metric coefficients. From expression \eqref{eq:sbs_surface}:
\begin{eqnarray}
\Bar{\rho}_{\rm kin} = e^{-v}\tilde{w}^2 \varphi^2   \approx \frac{e^{-v_\ast } \tilde{w}^2}{2 e^{2 \mathsf{m} _\ast z}+1} \label{eq:rho_kin_B} \, \\
\Bar{\rho}_{\rm st} = e^{-u} (\varphi')^2  \approx  \frac{4 e^{4 \mathsf{m} _\ast z}}{\left(2 e^{2 \mathsf{m} _\ast z}+1\right)^3} \label{eq:rho_st_B} \, \\
\Bar{\rho}_{\rm pot} = V \approx \frac{4 e^{4 \mathsf{m} _\ast z}}{\left(2 e^{2 \mathsf{m} _\ast z}+1\right)^3} \,, \label{eq:rho_pot_B}
\end{eqnarray}
with $z=\mathsf{r}-\mathsf{R^\ast}$ and $\Bar{\rho}_i=\Big( \frac{\mu^2\sigma^2_0}{2} \Big)^{-1}\rho_i$. In the thin wall limit one has $\Bar{\rho}_{\rm kin} \to H(-z)$, where $H$ is the Heaviside step function, while $\Bar{\rho}_{\rm st} \to \delta(z)$. From Eqns. \eqref{eq:u_struc}, \eqref{eq:v_struc} and the limiting cases, we can expect  that  $u$ will have a step-like behaviour at $R^\ast$, while $v$ will present a smoother transition. This type of behaviour was noted already in \cite{Friedberg:1986tq}. Solving Eq. \eqref{eq:u_struc} we find a complicated expression that we report in  App. \ref{app:analytic_uB}. For the configuration \rm{II} we show $u_{B}$ in Fig. \ref{fig:sbs_u_v_II} (Left).

\subsubsection{Energy balance} \label{sec:en_balance_an}

Like in the flat space-time case [Eq. \eqref{eq:qb_en_bal}], we can determine the inflection point in the case of SBSs by using energy balance arguments. We now need to account for the ``time'' dependence of the potential parameters as
\begin{equation}
\frac{d U_{\omega}}{dr}=\frac{\partial U_{\omega}}{\partial \phi} \phi'+\frac{\partial U_{\omega}}{\partial r} \,.
\end{equation}
One then finds
\begin{eqnarray} \label{eq:sbs_en_bal}
&& \frac{2}{\mu^2\sigma^2_0}\mathcal{E}(0)=  \\
&& \int^{\infty}_0 d\mathsf{r} \big[ \frac{2}{\mathsf{r}} (\varphi')^2-\frac{W'}{W} (\varphi')^2 +   \mathsf{m}'  \mathsf{m}  \varphi^2 (1-\varphi^2) - W'  W \varphi^2  \big]\,. \nonumber
\end{eqnarray}
Note that in the last expression, only the first term is present in the Minkowski limit \eqref{eq:qb_en_bal}, because $\mu W \to \omega$, $\mu \mathsf{m} \to \mu$ and $\omega, \mu$ do not run in ``time''. Like the first  one, the other terms are also dominant in the particular zones, as can be inferred from their form. The second and third terms are important in the boundary zone where the field interpolates between $\varphi_c$ and the exponential tail, while the fourth term receives important contributions both from the interior and the boundary zone. The leading order behaviour of all these terms is provided separately by
\begin{widetext}
\begin{eqnarray} \label{eq:sbs_en_bal_2_a}
A_{\mathcal{E}} &\equiv& \int^{\mathsf{R_>}}_{\mathsf{R_<}}d\mathsf{r} \frac{2}{\mathsf{r}} (\varphi')^2 \approx \frac{\mathsf{m}_\ast }{2 \mathsf{R^\ast}} \,,\\
B_{\mathcal{E}} &\equiv& -\int^{\mathsf{R_>}}_{\mathsf{R_<}} d\mathsf{r} \frac{W'}{W} (\varphi')^2 \approx -\frac{\mathsf{m}_\ast}{\mathsf{R^\ast}} \Big[
 \frac{1-\mathsf{m}^2_\ast}{4 }+\frac{1}{80} \mathsf{m}^2_\ast \Lambda^2 \mathsf{R^\ast}^2 \Big] \,, \\
C_{\mathcal{E}} &\equiv& \int^{\mathsf{R_>}}_{\mathsf{R_<}} d\mathsf{r}  \mathsf{m} ' \mathsf{m}   \varphi^2  (1-\varphi^2) \approx \frac{\mathsf{m}_\ast}{\mathsf{R^\ast}}   \Big[\frac{1-\mathsf{m}_\ast^2}{8} + \mathsf{m}_\ast^2 \Lambda^2 \mathsf{R^\ast}^2 \Big(\frac{1}{80} +\frac{1}{48} w^2 e^{-v_\ast} \Big)  \Big] \,, \\
D_{\mathcal{E} \, <} &\equiv& -\int^{\mathsf{R_<}}_0 d\mathsf{r} W' W \varphi^2  \approx   \frac{w^2 \varphi_c^2 }{2}  \big[1-  e^{u_<(\mathsf{R_<})-v_<(\mathsf{R_<})} \big] \,, \\
D_{\mathcal{E} \, B} &\equiv&  -\int^{\mathsf{R_>}}_{\mathsf{R_<}} d\mathsf{r} W' W \varphi^2  \approx  -2\frac{\mathsf{m}_\ast}{\mathsf{R^\ast}} w^2e^{-v_\ast} \Big[\frac{1}{2} \log \Big(\frac{3}{2}\Big) (1-\mathsf{m}^2_\ast)+ \frac{7 }{648}\mathsf{m}^2_\ast \Lambda ^2 \mathsf{R^\ast}^2 \Big] \,.  \label{eq:sbs_en_bal_2_d_s}
\end{eqnarray}
\end{widetext}
 The left hand side of Eq. \eqref{eq:sbs_en_bal} is therefore given as in Eq. \eqref{eq:qb_en_bal_expl}, i.e., neglecting exponentially suppressed terms, by
\begin{eqnarray} \label{eq:sbs_en_bal_3}
&&\Bar{\mathcal{E}}(0) \approx \frac{\left(\sqrt{3 w ^2+1}+2\right)}{27}  \left(3 w ^2+\sqrt{3 w^2+1}-1\right).
\end{eqnarray}

From the numerical results of Fig. \ref{fig:sbs_representative_lam}, it is clear that the effect of (strong) gravity is to introduce a new branch. As a result, for a subset of $\varphi_c \sim 1$ we have two different stable configurations [and possibly one more (un)stable one]  for the same $\varphi_c$. For the class of  configurations that contains \rm{I} and \rm{II}, we expect the Q-ball result $\mathsf{R^\ast} \sim 1/\tilde{w}^2$. For the most compact configurations,  from Eqns. \eqref{eq:u_asymp}, \eqref{eq:u_int}  and ignoring the jumping conditions, one naively gets
\beq
\mathsf{R^\ast} \sim \frac{\sqrt{12 \mathcal{C}_{\rm max}}}{\Lambda \tilde{w}} \sim \frac{2}{\Lambda \tilde{w}} \,.
\eeq
where we assumed in the last relation $\mathcal{C}_{\max} \approx \mathcal{C}_{\rm B+C}$. Although this expectation is too simplistic, it suggests the useful variable
\beq \label{eq:sbs_T_def}
T \equiv  \mathsf{R^\ast} \Lambda \tilde{w} \,.
\eeq
Because of the aforementioned degeneracy, $T(\tilde{w})$ is not a single-valued function. Instead, it is more useful to look for $\tilde{w}=\tilde{w}(T)$. The minimum of this curve, $\tilde{w}_\cup$, separates the Q-ball-like branch from the non-perturbative strong-gravity one, and provides the lowest $\tilde{w}$ for a given $\Lambda$. In flat space-time, one has $\tilde{w}_\cup=0$ $(\mathsf{R^\ast} \to \infty)$, so we expect $\tilde{w}_\cup \sim \mathcal{O}(\Lambda)$.

Approximating the complicated algebraic expressions in $\mathsf{R^\ast}$ we find (see App. \ref{app:analytic_T_full} for the details)
\beq \label{eq:sbs_w_T}
\tilde{w} \approx \Lambda  \frac{\sqrt{-\frac{T^4}{5}+6 T^2+36} \left(T^4+10 T^2+30\right)}{T \left(-T^4+30 T^2+180\right)}  \,.
\eeq
The minimum of this function corresponds to $\tilde{w}_\cup / \Lambda \approx 1.1$, while for $T \to 0$ we recover the Q-ball asymptotics $\tilde{w} \sim \Lambda  /T$. From there the approximate behaviour of $w$  follows:
\begin{eqnarray}
\frac{w}{\tilde{w}} \approx  && \Big(1+ \frac{5 T^2 \left(3 \tilde{w} ^2+4\right)-(2/3)T^4}{(2\sqrt{30})^2}\Big)^{-1/2} \times \nonumber \, \\
&& \Big(1+\frac{5 T^2 \left(3 \tilde{w}^2+8\right)+4 T^4}{(2\sqrt{30})^2}\Big)^{-1/2} \,, \label{eq:sbs_w_asymp_an}
\end{eqnarray}
which we show in Fig. \ref{fig:sbs_par_prost}. Our rough approximations have therefore given us a simple analytic description that predicts  the horizontal branching off of the $w-\varphi$ curve for any $\Lambda \ll 1$ (corresponding to the non-perturbative effect of strong-gravity), with  $\mathcal{C}$ getting close to LinEoS limit discussed in Sec. \ref{sec:sbs_maxC}.

\subsubsection{Finale: Semi-analytic solution} \label{sec:an_finale}

Instead of ignoring sub-leading terms, which led us to Eq. \eqref{eq:sbs_w_asymp_an}, we can also solve the full master (algebraic) Eqns. \eqref{eq:sbs_en_bal_2_a}-\eqref{eq:sbs_en_bal_3} numerically, using expression \eqref{eq:sbs_w_T} as a guess for each $T$. Now, we use $u_{B}$ when calculating $\mathsf{m}_{\ast}$ in Eqns. \eqref{eq:sbs_en_bal_2_a}-\eqref{eq:sbs_en_bal_2_d_s}, except in  $u_{B}$ itself [expression \eqref{app_eq_uB}], where  we iteratively take $\mathsf{m}_\ast \approx$  $\exp{(u_<(\mathsf{R}^\ast)/2)}$ . In contrast to the costly numerical solution of the boundary-value Einstein-Klein-Gordon system (App. \ref{AppEKG}), the semi-analytic approach provides a solution in a few seconds on a laptop computer.

\begin{figure*}[th]
\includegraphics[width=0.95\textwidth]{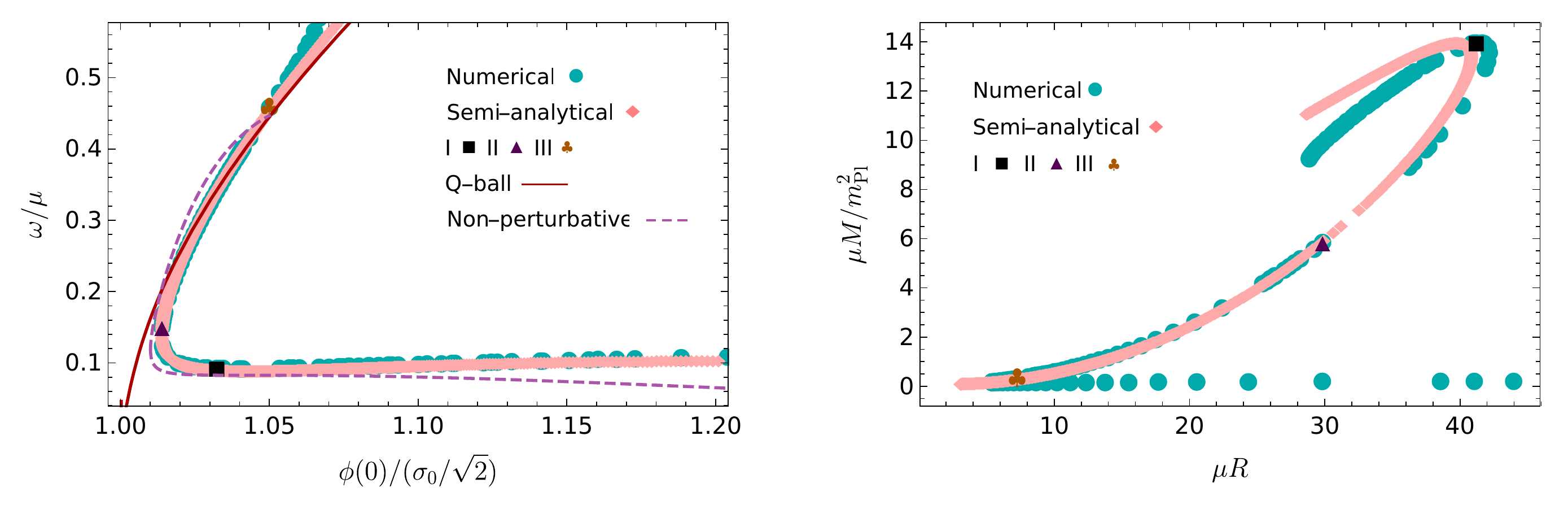}
\caption{(Left) Approximate analytic, semi-analytic (pink diamonds) and numerical calculations (cyan circles) of the
$w-\varphi$ behaviour for SBSs. The purple dashed line shows both branches from expression \eqref{eq:sbs_w_asymp_an}, while the dark red line represents the the Q-ball limit [expression \eqref{eq:qb_rad}, using expansion \eqref{eq:u_int}, \eqref{eq:v_int}]. (Right) Semi-analytic vs. numerical calculation of the $M-R$ curves. Both plots correspond to the benchmark scenario $\Lambda=0.186$. Three representative configurations from Table \ref{tab:1} are also shown.} 
\label{fig:sbs_par_prost} 
\end{figure*}

This procedure leads to a very good description of the configurations in the stable compact branch, as shown in Fig. \ref{fig:sbs_par_prost}. The semi-analytic results are in excellent agreement with the numerical ones for $\Bar{M}$, $v_\infty$, and consequently for $w$. For both $\mathsf{R^\ast}$ and $\mathsf{R}$, the agreement is almost perfect in the Q-ball-like strong-gravity branch (i.e. for  configurations similar to \rm{II}), while in the non-perturbative strong-gravity branch  (similar to \rm{III}), there is a systematic deviation, up to a few percent relative error. This error increases as  $\varphi$ increases, but note that this occurs mostly in the unstable branch, for which accurate approximations are
not of crucial importance. We elaborate more on the reason for the systematic error in App. \ref{app:analytic_error}.

Once the parameters are determined in the described way, one can use the expansions from the previous Subsections to reconstruct both the scalar and the gravitational field throughout the space-time, as well as the thermodynamic functions: density, pressure(s), speed of sound etc.

\section{Parameter space of Soliton boson stars} \label{sec:par_space}

In this Section, we will move away from the benchmark scenario of Sec. \ref{sec:gravitating}, where we only considered the compact stable branch of SBSs with the simple potential  \eqref{eq:V6qball} for $\Lambda \ll 1$. We will now consider the Planck limit $\Lambda \sim 1$ in Sec. \ref{sec:par_large_lam} and the low-compactness stable branch (for generic $\Lambda$) in Sec. \ref{sec:par_dilute}. We will then  adopt potentials with multiple degenerate vacua in Sec. \ref{ref:par_sextic}, ones with a false vacuum instead of a degenerate one  in Section
\ref{sec:par_v6}, and we will test the robustness of
our conclusions in Sec. \ref{sec:par_v4}, by considering a non-polynomial effective potential.

\subsection{Planck scale regime} \label{sec:par_large_lam}

To understand the qualitative impact of a large ``control parameter'' $\Lambda$, let us assume that the Q-ball description is valid up to $\mathcal{C}_{\rm{B+C}}$ i.e. $\mathcal{C}_{\rm{max}} \sim$ $\frac{\Lambda^2}{16\pi} \varphi^2_c \mathsf{R}^2 w^2$ $\propto \Lambda^2 / w_{\rm{min}}^2$ $<\mathcal{C}_{\rm{B+C}}$. As $\Lambda$ increases, in order for  $\mathcal{C}_{\rm{max}}$
to asymptote to $C_{\rm{B+C}}$ in the thin wall regime, $w_{\rm{min}}$ has to increase. However, as $w_{\rm{min}}$ increases, the thin wall regime is superseded by the thick wall one, and  when $w_{\rm{min}} \simeq  w_{\rm{Qb-s}} $ [cf. Eq. \eqref{eq:qb_w_max}] the stable branch inherited from the flat space-time limit disappears.

\begin{figure*}[th]
\includegraphics[width=\textwidth]{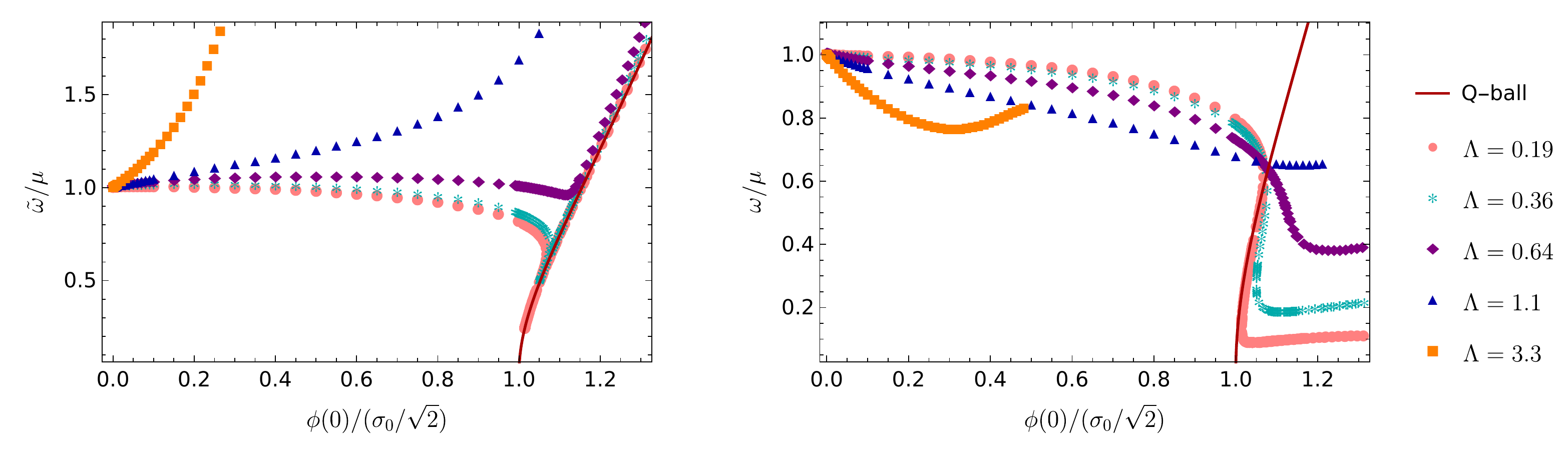}
\caption{(Left) Frequency vs central scalar field value in
$\tilde{v}(0)=0$ gauge and (Right) in the $v(\infty)=0$ gauge, for SBSs with different values of $\Lambda$, along with the (gauge-independent) analytical result for Q-balls \eqref{eq:qb_central}. }
\label{fig:parsp_fiwwT}
\end{figure*}

\begin{figure}
\centering
\includegraphics[width=0.42\textwidth]{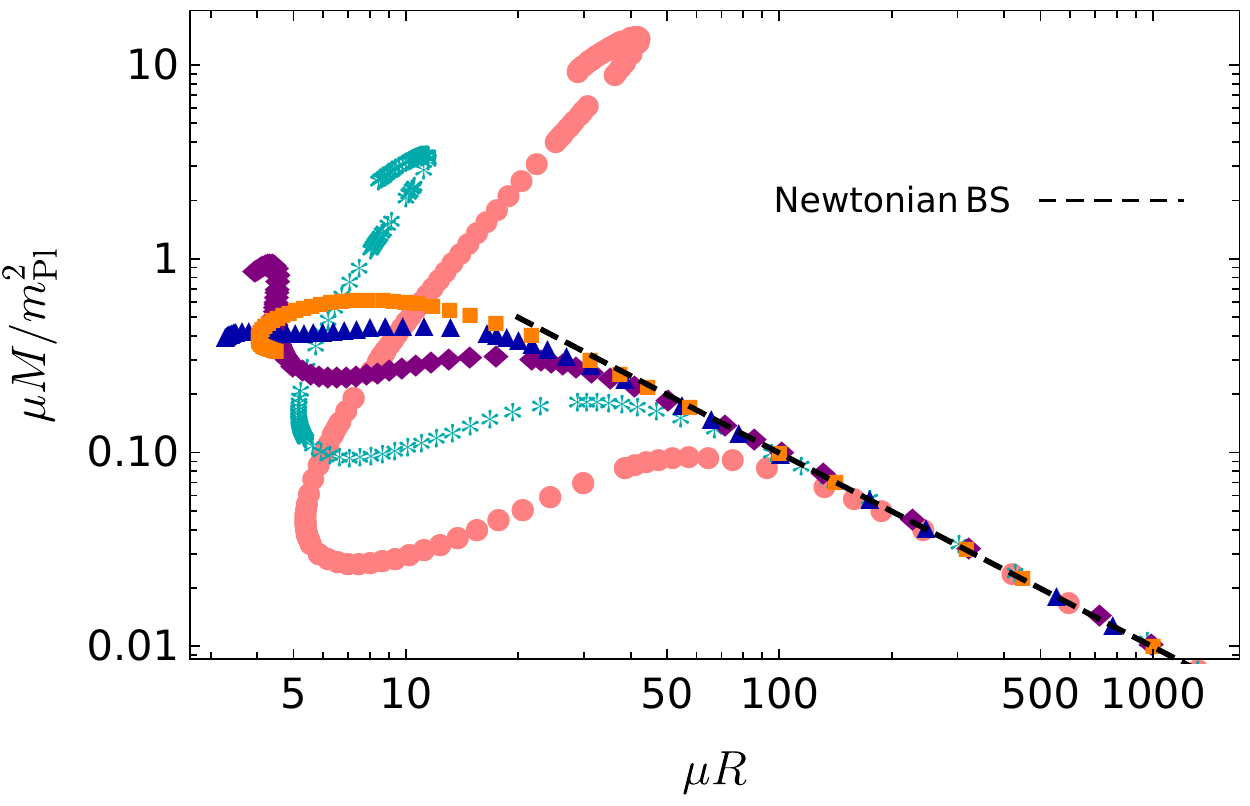}
\caption{Mass-radius relation for SBSs with different values of $\Lambda$. The dots represent numerical results, while the dashed black line represents the analytic result for NBS \eqref{eq:wf_mbs_MR}. The shapes and colours of dots are the same as in Fig. \ref{fig:parsp_fiwwT}.}
\label{fig:parsp_MR}
\end{figure}

\begin{figure}
\centering
\includegraphics[width=0.42\textwidth]{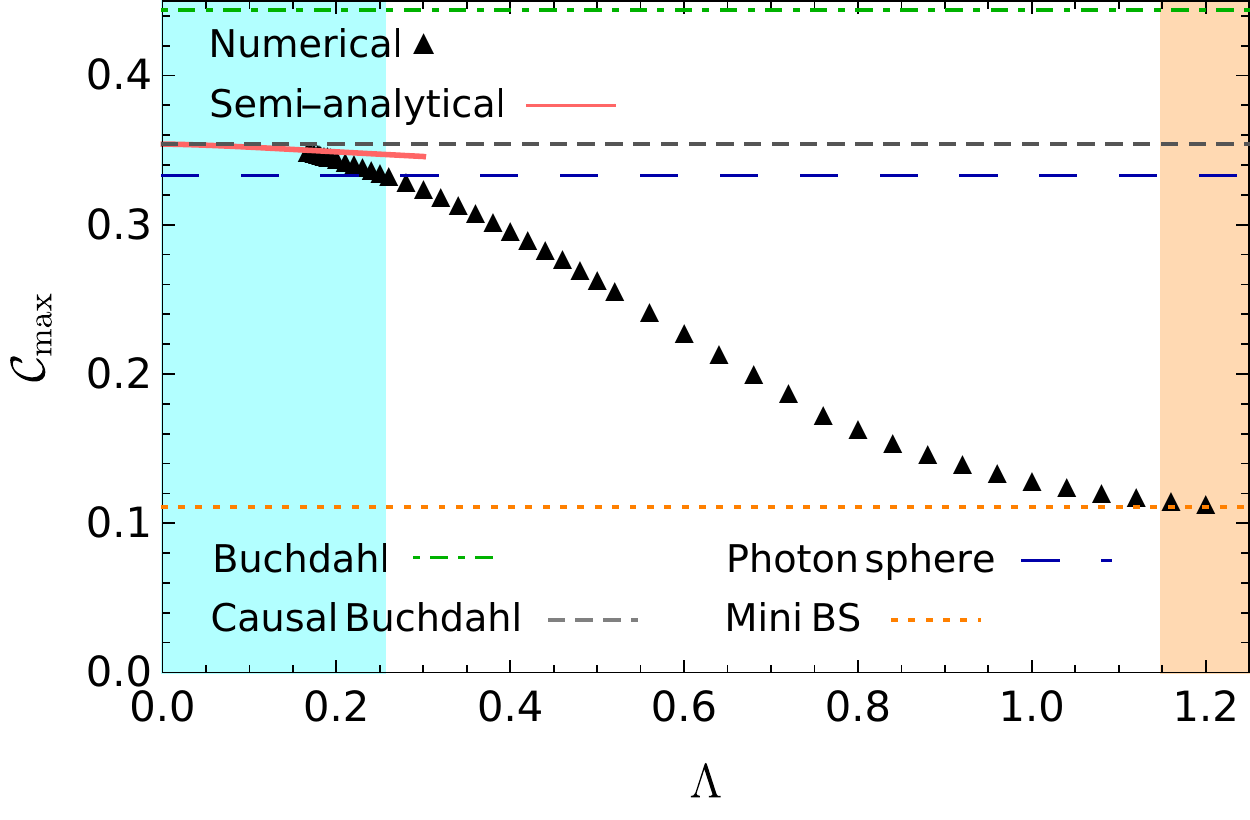}
\caption{Maximum compactness of SBSs for given $\Lambda$. We have also indicated the Buchdahl bound $\mathcal{C}_{\rm B}=0.44$, the causal Buchdahl bound $\mathcal{C}_{\rm B+C}=0.354$, the condition for the photon sphere $\mathcal{C}=0.33$ and the maximal compactness for MBSs $\mathcal{C}=0.11$. In the ultra-compact domain (cyan region), the numerical results are supplemented with semi-analytical ones. Note that for $\Lambda =0$: $\mathcal{C}_{\rm{max}} \to \infty$ (Q-ball limit). The orange region represents the MBS part of the parameter space and extends to $\Lambda \to \infty$.}
\label{fig:parsp_MC}
\end{figure}

We have presented numerical results for $\tilde{w} \,, w$ as functions of $\Lambda$ in Fig. \ref{fig:parsp_fiwwT}. Note that the dip in the $w(\varphi_c)$ curve increases with $\Lambda$. This corresponds to the growth of the height of the horizontal asymptote of the $w(\varphi_c)$ and hence minimal possible $w$ as argued above. Consequently the Planck scale regime leads to \textit{less} compact configurations (Fig. \ref{fig:parsp_MC}). As strong deviations from the Q-ball description occur in the compact stable branch at $\Lambda \simeq 0.4$, we cannot make quantitative predictions for when the compact stable branch will disappear, but numerically we find that this occurs at $\Lambda \simeq 1.1$. The mass-radius relation is shown in Fig.  \ref{fig:parsp_MR}, while Fig. \ref{fig:parsp_MC}  represents how the maximum achievable compactness changes\footnote{It is interesting to note that scalar stars in Horndeski's theory have been recently constructed and present very similar a behaviour for $\mathcal{C}$ (c.f. Fig. 10 in \cite{Barranco:2021auj}). However, we are not aware of a simple mapping between SBSs in GR and this kind of configurations.} from $C_{\rm B+C}$ to $C_{\rm MBS} \approx 0.11$ over $\Lambda \sim 0.1-1$ range.

\subsection{Low compactness stable branch} \label{sec:par_dilute}

The low compactness stable branch of SBSs is supported by  quantum pressure,
and as a result in this limit SBSs behave as MBSs. The exact MBS limit of SBSs is $\Lambda \to \infty$, and the appropriate field parametrization is
\begin{equation}
 \varphi_{\rm MBS}=\phi/M_{\rm Pl}  \,,
\end{equation}
as there are no scales on which the structure equations depend, except for the Planck scale. These configurations were thoroughly studied  in the original paper by Kaup \cite{Kaup:1968zz}  and are reviewed in \cite{Liebling:2012fv,Schunck:2003kk}. The numerical procedure to obtain MBSs  is analogous to the SBS case (App. \ref{AppEKG}).

In the SBS setting, at $\Lambda \simeq 1.1$ the compact stable branch of SBS vanishes, and only the MBS one is left. For small $\Lambda$, the low compactness stable branch is replaced
by the unstable Q-ball branch in the weak field regime. As the control parameter increases from $\Lambda \simeq 0.7$, more and more configurations in the low compactness stable branch develop \textit{higher} compactness, up to the MBS limit of $C_{\rm MBS} \approx 0.11$. The low compactness branch presents different behaviour than the compact one, where the Planck scale regime leads to lower compactness. Note that the highest compactness is achieved in the unstable branch, while in the stable one $C^{(s)}_{\rm MBS} \approx 0.08$.

In the weak-field approximation of MBSs, the  Einstein-Klein-Gordon system reduces to the Schr\"odinger-Poisson   system (Newtonian boson stars) \cite{Seidel:1990jh, Guzman:2003kt, Liebling:2012fv, Kling:2017mif, Boskovic:2018rub, Annulli:2020lyc}
\begin{eqnarray}
e \phi = - \frac{1}{2\mu} \nabla^2 \phi + \mu \phi \Omega \,,  \\
\nabla^2 \Omega = \frac{1}{2} M^{-2}_{\rm Pl} \mu^2 \phi^2 \,,
\end{eqnarray}
where $e^v \approx 1+ 2 \Omega$,  $\omega \approx \mu + e$ and $\Omega \ll 1$, $e \ll \mu$. Like in the general discussion, the scalar mass can be factored out, and in addition  system admits a scaling symmetry
\begin{eqnarray}
\phi \to k^2 \phi \,, e \to k^2 e \,, r \to r/k \,, \Omega \to k^2 \Omega \,.
\end{eqnarray}
This allows for a universal description of these objects,  and for performing the numerical integration  only once (for the analytic solution see \cite{Kling:2017mif, Kling:2017hjm}).

Fixing the scale with $-k^2=e/2$, we can find several useful relations between the macroscopic parameters, which will be compared with the relativistic numerical results, e.g.
\begin{eqnarray}
\Bar{M} = \frac{\beta Z}{\Bar{R}} \,, \label{eq:wf_mbs_MR} \\
\Bar{M} = \sqrt{2\beta^2 \Big(1-w \Big)} \,,  \label{eq:wf_mbs_Mw}  \\
\varphi_c=\frac{2s_0}{\Lambda} \frac{1}{\sqrt{w}} \Big(1 - w \Big)   \label{eq:wf_mbs_fiw} \,,
\end{eqnarray}
where  $s_{0} = 1.022$ and $\beta = 1.753$ \cite{Kling:2017mif} and we find the scale-invariant radius (that encloses 99\% of the BS mass) to be $Z=5.6741$.

In Fig. \ref{fig:parsp_MR}, we see that the Newtonian boson star scaling gives a good description of SBS configurations in the $\varphi_c \to 0$ limit.

\subsection{Cosine potential} \label{ref:par_sextic}

In axionic physics, a cosine potential $V \sim \cos{(a/f_a)}$, where $a$ is the axion field and $f_a$ is a decay constant,  is often considered. This potential arises from non-perturbative effects that generate small masses for the (initially massless) Goldstone boson associated with the spontaneous breaking of the Peccei-Quinn symmetry \cite{Marsh:2016rep,Hui:2016ltb,Helfer:2017a}. In the strong gravity context, an axion potential can produce non-trivial effects on the stability of axion stars  \cite{Helfer:2017a}.

Inspired by  axion star solutions with the axion potential, some authors have considered boson star models with  similar potentials \cite{Guerra:2019srj,Delgado:2020udb,Siemonsen:2020hcg}. Note that in the absence of
beyond standard model physics
that could motivate such potentials for  complex scalars with $U(1)$ symmetry, one should consider these models only as proxies to understand (pseudo-real) axion stars (and only if different minima are physically sensible). As these potentials develop multiple minima and having in mind the Taylor expansion of the $\cos$ function, our previous discussion would imply that ``axion boson stars'' would periodically replicate SBSs for the different values of $\Lambda$ corresponding to different minima, up to the Planckian threshold.

\begin{figure*}[th]
\begin{tabular}{cc}
\includegraphics[width=.42\textwidth]{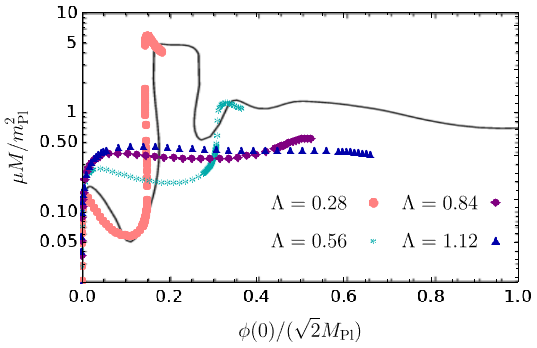}
\qquad
\includegraphics[width=.43\textwidth]{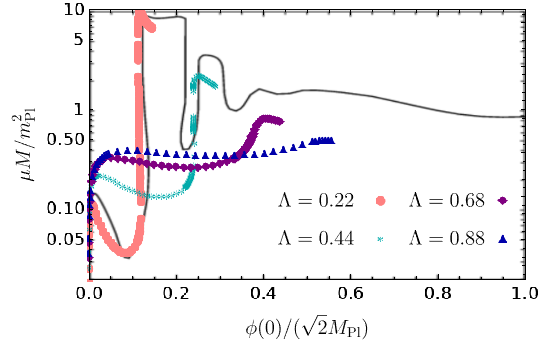}
\end{tabular}
\caption{Mass-central field dependence for the cosine potential  \eqref{eq:cos_pot}. The black line represents numerical results from \cite{Guerra:2019srj}, while dots correspond to  a set of SBSs with $\Lambda_n$ from Eq. \eqref{eq:cos_lambda} and $n=1,2,3,4$. The left panel represents cosine potential boson stars with $f_a/m_{\rm Pl}=10^{-2.2}$, while the right one assumes $f_a/m_{\rm Pl}=10^{-2.3}$, with the other parameters set to the values used in \cite{Guerra:2019srj}. Note the different rescaling of the scalar with respect to the rest of this work, in accordance with the conventions of \cite{Guerra:2019srj}.}
    \label{fig:parsp_cos}
\end{figure*}

For concreteness we will consider a specific form of the potential, from \cite{Guerra:2019srj}:
\begin{equation} \label{eq:cos_pot}
V=\frac{2\mu^2 f^2_a}{B} \Big[1- \sqrt{1- 4 B \sin^2 \Big(\frac{|\Phi|}{2 f_{\rm a}} \Big)} \Big] \,,
\end{equation}
where $B$ is a model dependent constant [taken to be $B \approx 0.22$ in \cite{Guerra:2019srj}]. The
minimum of  potential \eqref{eq:cos_pot} occurs at
\begin{equation} \label{eq:cos_min}
\phi_{\rm min}=f_a 2 n \pi \,, n \in \mathbb{N} \,.
\end{equation}
This gives an $n$-dependent $\Lambda$ scale
\begin{equation} \label{eq:cos_lambda}
\Lambda_{n}=\frac{f_a}{m_{\rm Pl}} 2n \pi \sqrt{16 \pi} \,, n \in \mathbb{N} \,.
\end{equation}
In Fig. \ref{fig:parsp_cos}, we have compared numerical results from \cite{Guerra:2019srj} with a set of SBSs specified by the potential \eqref{eq:V6qball} and the control parameter \eqref{eq:cos_lambda}. It is clear that the periodic features for the ``axion boson stars'' occur for the appropriate field values given by \eqref{eq:cos_min}. The $\Bar{M}-\phi(0)$ plot for the first minimum is in excellent quantitative agreement with the SBS results. The agreement, however, is not perfect as $\cos$ can only locally be approximated with the sextic polynomial. As we progress in $n$, the agreement worsens. This should come as no surprise, because in the analogue perspective an  axion boson star ``particle'' has to go through an effective potential that has several peaks and troughs, unlike in the SBS case. Finally, a sufficiently large $n$ field is in the Planck scale regime, and the final unstable branch is reached, like in the SBS case (Sec. \ref{sec:par_large_lam}).

The cosine potential example is illustrative also for the following reason: the sextic potential is non-renormalizable
and thus not valid for arbitrary field values  (as one needs to be within the limit of validity of the effective field theory). Close to this limit, higher-dimensional operators become relevant. The cosine potential illustrates that such terms do not modify qualitatively the macroscopic behaviour, as long as degenerate vacua (and even false ones, as we will argue next)  are present.

\subsection{General sextic potential} \label{sec:par_v6}

The case of degenerate vacua is somewhat special, while a more generic scenario would allow for a non-zero false vacuum:
\begin{eqnarray}  \label{eq:v6_1}
V = \mu^2 |\Phi|^2 -\beta |\Phi|^4 + \xi |\Phi|^6 \,, \\
\beta>0 \,,\xi>0 \,. \nonumber
\end{eqnarray}
There are two useful reparametrizations of this potential. The first one, used in \cite{Heeck:2020bau}, parametrizes the potential as a deviation from the degenerate vacuum case:
\begin{eqnarray}
&& V_6 = \phi^2_0 \Big[ (\mu^2-\omega^2_0) \varphi^2 (1-\varphi^2)^2 + \omega^2_0 \varphi^2 \Big]
\,, \label{eq:v6_2} \\
&&  \varphi= \frac{|\Phi|}{\phi_0} \,,\, \phi_0= \sqrt{\frac{\beta}{2\xi}} \,,\,  w_0= \frac{\omega_0}{\mu}=\sqrt{1- \frac{\beta^2}{4 \xi \mu^2}} \nonumber
\end{eqnarray}
The parameter choices $\beta=4 \mu^2 /\sigma_0^2$, $\xi=4 \mu^2 /\sigma_0^4$ imply  $\omega^2_0=0$ and $\phi_0=\sigma_0/\sqrt{2}$ and  reproduce the  potential \eqref{eq:V6qball} i.e. the benchmark scenario of this work, while $\omega_0$ parametrizes deviation from the simplest potential \eqref{eq:V6qball}. Another useful approach is to relate $\omega_0$  to the ratio between the potential barrier $\phi_{\rm B}:$ $dV/d\phi|_{\rm B}=0$ \,, $d^2V/d\phi^2 |_{\rm B}<0$ and the non-trivial minimum (false vacuum)   $\phi_{\rm  F}$:
\begin{eqnarray}\label{eq:V6x}
V_6 &=& \frac{\mu^3}{6 \sqrt{3\xi}  }\frac{\varphi_x^2}{x^3}\Big(6 x^2+ \left(-3 x^2-3\right) \varphi_x^2+2 \varphi_x^4 \Big) \,, \\
\varphi_x &\equiv& \frac{\phi}{\phi_{\rm B}} \,, \phi_{\rm F/B} = \frac{1}{\sqrt{3\xi}} \sqrt{\beta \pm \sqrt{\beta^2-3\xi \mu^2}} \,, \nonumber \\
x &\equiv& \frac{\phi_{\rm B}}{\phi_{\rm F}} = \sqrt{\frac{2-\sqrt{1-3 \mathsf{w}_0^2}}{2+\sqrt{1-3 \mathsf{w}_0^2}}} \,, \mathsf{w}_0=\frac{\omega_0}{\sqrt{\mu^2-\omega_0^2}} \nonumber \,.
\end{eqnarray}
In this parameterization, the limits $x=1/\sqrt{3}$ ($\mathsf{w}_0=0$) and $x=1$ ($\mathsf{w}_0=1/\sqrt{3}$) interpolate between the degenerate vacua case and the scenario where the potential develops an exact stationary inflection point. Allowing for $x<1/\sqrt{3}$ (and hence imaginary $w_0$) makes $\varphi_{\rm F}$ a true vacuum, but we will not consider that scenario in this work\footnote{See \cite{Sakai:2007ft} for the flat space-time case and  \cite{Tamaki:2011zza} for the gravitating case.}. In principle, Coleman's (necessary) stability criterion (c.f. Sec. \ref{sec:qb}) $w_0 \leq w < 1$ allows  even for the cases where the second minimum  disappears $\mathsf{w}_0>1/\sqrt{3}$  $(w_0>1/2)$ and the parametrization given in Eq. \eqref{eq:V6x} is not applicable.
However, such configurations are in the deep thick wall regime, as we will argue below. Thus,
defining the class of non-topological solitons that we have examined in this work by the presence of the false/degenerate vacuum in the potential is not completely rigorous. However, not only does that describe the largest part of the parameter space, but also the region where
the phenomenology of these objects  differs most significantly from ``normal matter''.

We can define (as in \cite{Heeck:2020bau})
\begin{equation} \label{eq:v6_kappa}
\varkappa^2=\frac{w^2-w^2_0}{1-w_0^2}\,,
\end{equation}
so that the Minkowski Klein-Gordon equation has the same form as in Sec. \ref{sec:qb_simp} if one substitutes $w \to \varkappa$ and $\mathsf{r} \to \sqrt{\mu^2-\omega_0^2}r$. One can thus use both analytic and numerical results for the scalar profile of Q-balls with the simplest potential. The macroscopic properties $R, M, Q...$, however, depend explicitly on $w_0$ (see \cite{Heeck:2020bau} for the relevant expressions). As $w_0$ increases from $0$ to $1$, the length scale of the boundary $\propto (1-w^2_0)^{-1/2}$ increases  even if $\varkappa \ll 1$. Thus, as the false vacuum departs from $0$, the thick-wall regime increasingly  dominates the Q-ball behaviour.

In  curved spacetime,  the Klein-Gordon equation is not invariant under the above reparametrization\footnote{This is expected from the equivalence principle: in Minkowski space it is enough to know the energy difference between the two vacua, while in GR information about both vacua is needed.}. Instead, the system \eqref{eq:u_struc} -  \eqref{eq:SBS_KG} can be formulated as:
\begin{eqnarray}
&& \frac{1}{\mathsf{r}^2}\l(\mathsf{r}\,e^{-u}\r)' -\frac{1}{\mathsf{r}^2}= -\frac{\Lambda^2}{2} \times \label{eq:v6_u} \\
&& \Big[e^{-v}\varkappa^2 \varphi^2+e^{-u} (\varphi')^2 +\varphi^2 (1-\varphi^2)^2 + \mathsf{w}_0^2 \varphi^2 (1+e^{-v})\Big]\,, \nonumber \\
&& e^{-u}\l(\frac{v'}{\mathsf{r}}+\frac{1}{\mathsf{r}^2}\r)-\frac{1}{\mathsf{r}^2}=\frac{\Lambda^2}{2} \times  \label{eq:v6_v} \\
&&  \Big[e^{-v}\varkappa^2 \varphi^2+e^{-u} (\varphi')^2 -\varphi^2 (1-\varphi^2)^2 - \mathsf{w}_0^2 \varphi^2 (1-e^{-v})\Big]\,, \nonumber \\
&&\varphi''+\l(\frac{2}{\mathsf{r}}+\frac{v'-u'}{2}\r)\varphi'= \label{eq:v6_KG} \\
&& e^u\l[(1-4\varphi^2+3\varphi^4)-\varkappa^2 e^{-v}+ \mathsf{w}_0^2(1-e^{-v}) \r]\varphi\,, \nonumber
\end{eqnarray}
where
\begin{equation}
\mathsf{r} = \sqrt{1-w_0^2} \, \mu r \,, \Bar{m}(\mathsf{r})=\frac{\sqrt{1-w_0^2} \mu m(r)}{m^2_{\rm Pl}} \,, \Lambda= \frac{\sqrt{2} \phi_0}{M_{\rm Pl}} \nonumber
\end{equation}
and the other conventions from Eqs. \eqref{eq:v6_2},  \eqref{eq:V6x} and \eqref{eq:v6_kappa} apply, while the prime $'$ denotes spatial derivatives with respect to $\mathsf{r}$. In the Minkowski limit, one has $u \to 0, v \to 0$, and the explicit dependence on $\mathsf{w}_0^2$ disappears in the Klein-Gordon equation.

\begin{figure*}[th]
\includegraphics[width=\textwidth]{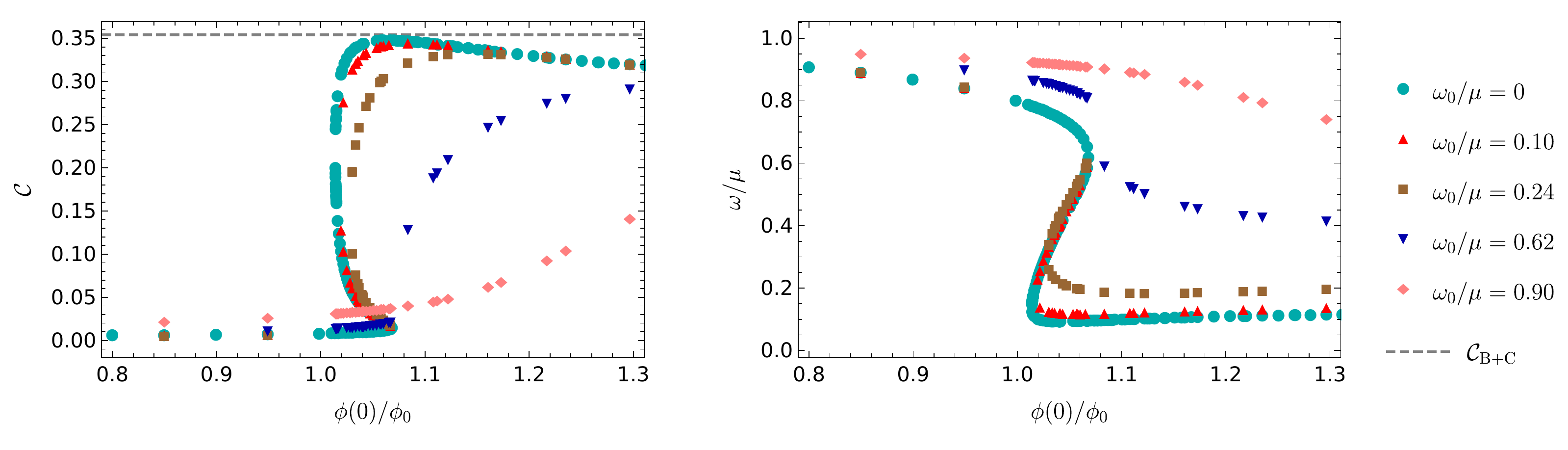}
\caption{Compactness (left) and scalar frequency (right) of SBSs for the general sextic potential \eqref{eq:v6_1}, in the  $v(\infty)=0$ gauge,  as a function of the central field. Various values of the parameter $\omega_0$ that describes the deviation from the degenerate vacuum case are considered, and we set $\Lambda=0.186$, comparing also with the vanila case $\omega_0/\mu=0$.} \label{fig:v6_C_fi_w}
\end{figure*}

Self-gravitating configurations with the potential  \eqref{eq:v6_1} have been considered for particular values of the coefficients in \cite{Kleihaus:2005me,Tamaki:2011zza, Kleihaus:2011sx, Siemonsen:2020hcg}. The parameterization outlined here allows us to perform a systematic exploration of the parameter space, by varying $w_0$ from $0$ to $1$ in discrete steps using the same numerical approach as in the rest of this work (App. \ref{AppEKG}). In Fig. \ref{fig:v6_C_fi_w} (left) we show how the compactness decreases (for fixed $\Lambda$) from the most compact configurations $w_0=0$  to $w_0 \to 1$.  This  is represented in the $w - \varphi$ parameter space by the growth of the height of the horizontal asymptote,
or by that of the tipping point of the two branches in the $\tilde{\omega} - \varphi$ parameter space, as demonstrated in Fig. \ref{fig:v6_C_fi_w} (right).

The interpretation of these results is straightforward: increasing the height of the false vacuum implies thicker walls and hence larger minimal frequency/smaller maximal mass and compactness. The picture outlined in Sec. \ref{sec:sbs_an}, where the analogue particle in the $\Tilde{v}(0)=0$ gauge does not initially feel the presence of gravity, is valid also  for the general sextic potential. Henceforth, the curve $\tilde{w} - \varphi$ in the thin-wall regime is given by Eq. \eqref{eq:qb_central} (with $w \to \varkappa$):
\begin{equation}
\Tilde{w}^2 = (1 - 4 \varphi^2_c
 + 3 \varphi^4_c)(1-w^2_0)+w^2_0 \,.
\end{equation}

The compactness dependence on $w_0$ can also be understood in terms of the LinEoS. Using an arguments analogous to those that lead to Eq. \eqref{eq:linEoS_cs_qb} one gets, for the general sextic potential, the following estimate of the speed of sound
\begin{eqnarray}
(c^2_s)_a = \frac{\varphi_c^2 \left(w_0^2 \left(3 \varphi_c^2-4\right)+6 \varphi_c^2-4\right)}{2+4 \left(w_0^2-3\right) \varphi_c^2 - 3 \left(w_0^2-4\right) \varphi_c^4} \label{eq:cs_sextic} \,. 
\end{eqnarray}
This equation predicts that even for $\varphi_c = 1$, the speed of sound will be subluminal when $w_0 \neq 0$. For example, the compactness of the maximum mass  configuration with $w_0=0.24$ is numerically found to be $\mathcal{C}=0.321$, while Eqns. \eqref{eq:cs_sextic} and \eqref{eq:LinEos_C_w} predict a similar value for the corresponding $\varphi_c=1.083$: $\mathcal{C}_{\rm{LinEoS}}=0.326$.

In agreement with the discussion in Sec. \ref{sec:par_dilute}, for $\Lambda \geqsim 1$ the configurations exhibit a MBS-like behaviour, irrespective of the value of $w_0$.

\subsection{Non-polynomial quartic potential} \label{sec:par_v4}

As one last departure from the benchmark scenario of this work, we will now consider  a real scalar field $\phi$ with the renormalizable potential
\begin{eqnarray} \label{eq:V4}
V_4(\phi) = \mu^2 \phi^2 - g \phi^3 + \lambda \phi^4 \,,  \\
g>0 \,, \lambda>0 \,.  \nonumber
\end{eqnarray}
Q-balls with this effective potential can form in the presence of other fields \cite{Kusenko:1997zq,Postma:2001ea,FSS}, or we can consider this scenario as a proxy for a pseudo-soliton composed of real scalars. Formally (and in line with the rest of this work) we will take this model to originate from the non-polynomial potential of a $U(1)$ complex scalar
\begin{equation} \label{eq:V4cs}
V_4(|\Phi|) = \mu^2 |\Phi|^2 -g (|\Phi|^2)^{3/2} + \lambda |\Phi|^4\,,
\end{equation}
with the ansatz \eqref{eq:qb_field} [$\Phi=\phi(r) e^{-i\omega t}$] giving a real scalar $\phi$ in Eq. \eqref{eq:V4}.

Like for the generic sextic potential of Sec. \ref{sec:par_v6},  the potential of Eq. \eqref{eq:V4} admits false and degenerate vacua, and can be parameterized as a deviation from the degenerate case
\begin{eqnarray}
&& V_4 = \phi^2_0 \Big[ (\mu^2-\omega^2_0) \varphi^2 (1-\varphi)^2 + \omega^2_0 \varphi^2 \Big]
\,, \label{eq:v4_2} \\
&&  \varphi= \frac{|\Phi|}{\phi_0} \,,\, \phi_0=\frac{g}{2\lambda} \,,\, w_0=\frac{\omega_0}{\mu}= \sqrt{1- \frac{g^2}{4\lambda \mu^2}}\,, \nonumber
\end{eqnarray}
or parameterizing the two vacua
\begin{eqnarray}\label{eq:V4x}
V_4 &=& \frac{\lambda \phi^4_{\rm F}}{3}  \varphi_x^2 \left(6 x  -4 (x+1) \varphi_x +3 \varphi_x^2\right) \,, \\
 \varphi_x &\equiv& \frac{\phi}{\phi_{\rm B}} \,, \phi_{\rm F/B} = \frac{3g \pm \sqrt{9 g^2-32 \lambda  \mu^2}}{8 \lambda } \,, \frac{1}{\mu^2 \phi^2_{\rm F}}\frac{\lambda \phi^4_{\rm F}}{3} = \frac{1}{6x} \,, \nonumber \\
x &\equiv& \frac{\phi_{\rm B}}{\phi_{\rm F}} =\frac{3-\sqrt{1-8 \mathsf{w}_0^2}}{3+\sqrt{1-8 \mathsf{w}_0^2}} \,, \mathsf{w}_0=\frac{\omega_0}{\sqrt{\mu^2-\omega_0^2}} \nonumber \,.
\end{eqnarray}

In contrast to the $V_6$ case in flat spacetime, all configurations are  (classically) stable provided that\footnote{Quantum effects can influence stability in part of the parameter space for small Q-balls \cite{Postma:2001ea}. Note that
stable solutions (under small perturbations)
can exist also for $w^2_0 < 0$   \cite{Paccetti:2001uh,Sakai:2007ft}.}  $w^2_0 \geq 0$ \cite{Kusenko:1997ad,Paccetti:2001uh,Sakai:2007ft}. Otherwise, the discussion is similar to the generic $V_6$ case: for $1 \gg w^2 \sim w^2_0$ Q-balls are in the thin wall regime, while for  $w \sim 1$  they are in the thick wall one.

The gravitating case for this potential was considered in \cite{Tamaki:2010zz}, but  only for one value of (in our parametrization) $w_0  \approx 0.4 $, which is in the intermediate thick wall regime. Expecting a similar phenomenology with respect to $w_0 \neq 0$ as in Sec.
\ref{sec:par_v6}, we have focused only on the $w_0 = 0 $ case (degenerate vacua), noting that the above parametrization allows for a straightforward systematic exploration with respect to the height of the false vacuum.

\begin{figure}
\centering
\includegraphics[width=0.42\textwidth]{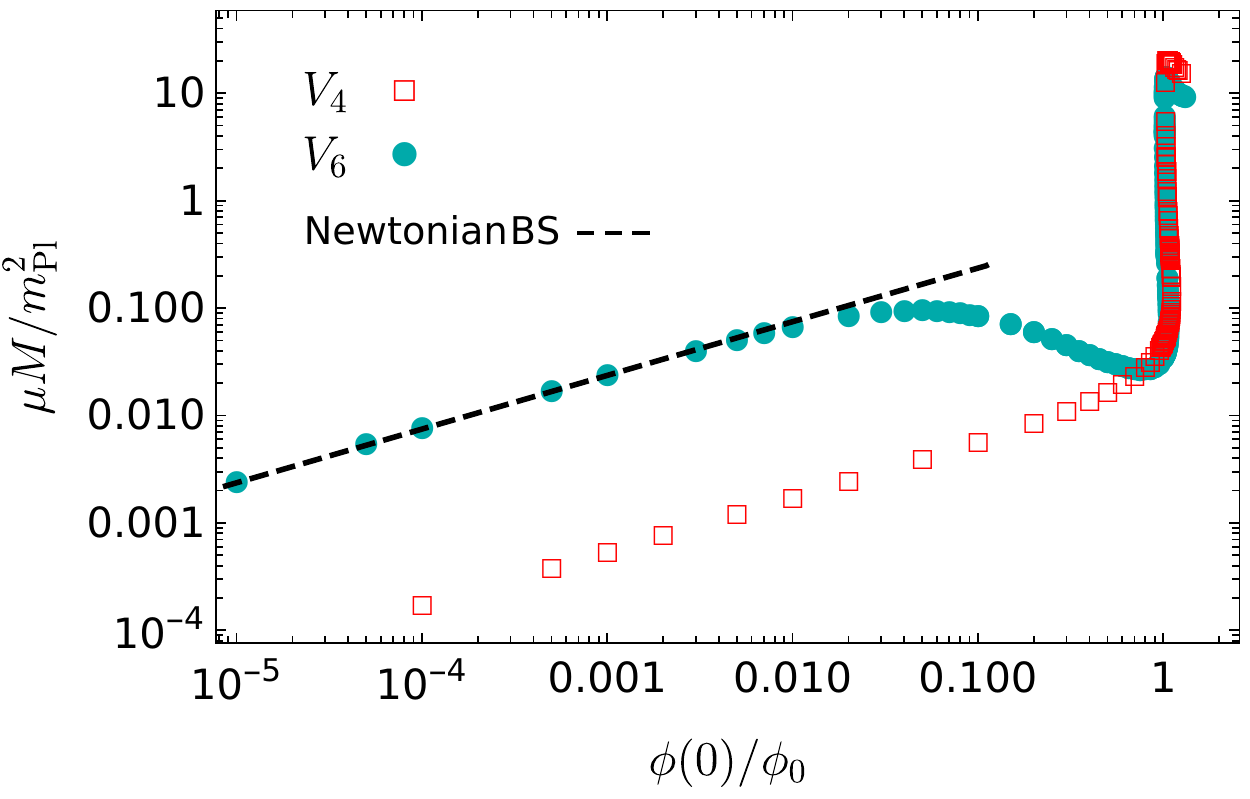}
\caption{Mass vs central field of SBSs in the case of the quartic potential \eqref{eq:V4}, compared with the benchmark case \eqref{eq:V6qball}  for the same control parameter $\Lambda=0.186$. We also show the NBS scaling  \eqref{eq:wf_mbs_fiw} (black, dashed).}
\label{fig:v4_fi_M}
\end{figure}

\begin{figure}
\centering
\includegraphics[width=0.42\textwidth]{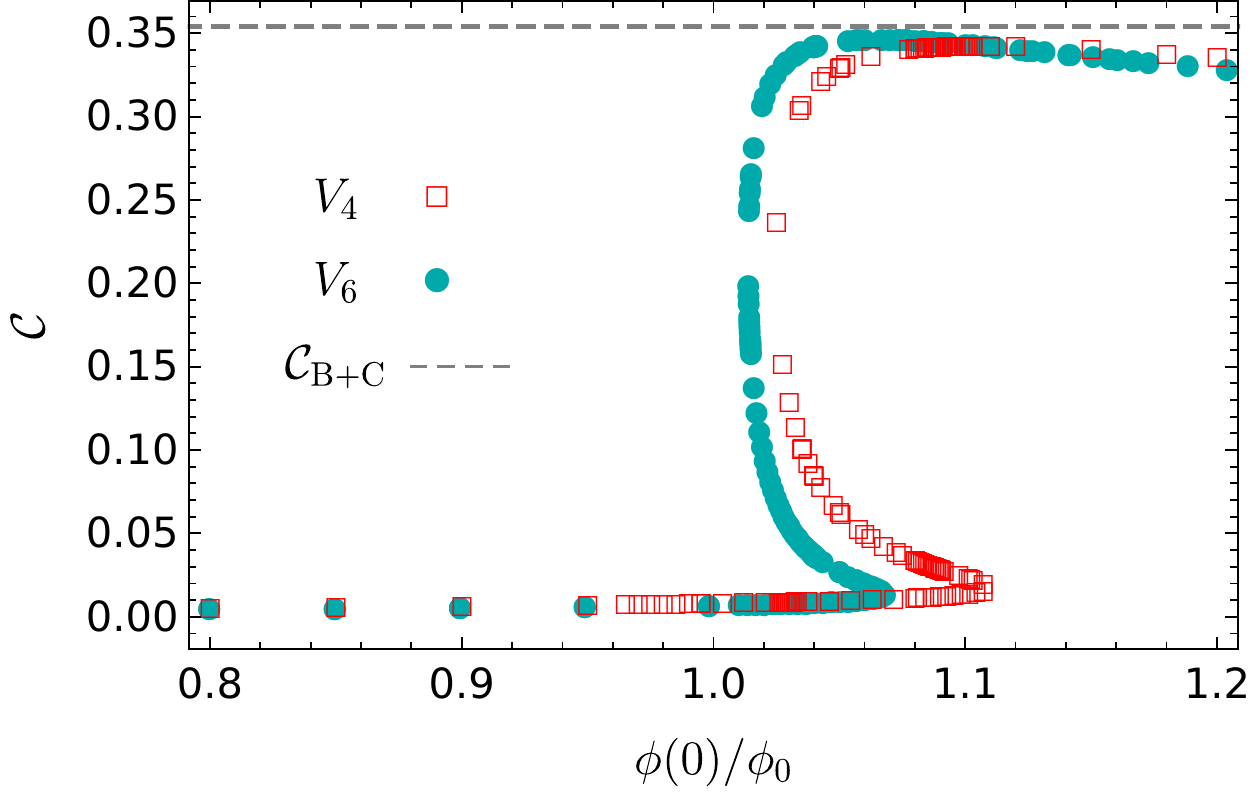}
\caption{Compactness vs central field of SBSs in the case of the quartic potential \eqref{eq:V4}, compared with the benchmark case \eqref{eq:V6qball}  for the same control parameter $\Lambda=0.186$. The causal Buchdahl bound $\mathcal{C}_{\rm B+C}$ is indicated as a gray and dashed line.}
\label{fig:v4_fi_C}
\end{figure}

In the thin-wall gravitating limit,
the relations among the macroscopic parameters are very similar to those of SBSs with $V_6$ [Eq. \eqref{eq:V6qball}], as can be seen in Fig.  \ref{fig:v4_fi_M} for the mass and Fig. \ref{fig:v4_fi_C} for the compactnesses. This should come as no surprise as the presence of the non-trivial vacuum renders the equation of state linear in the thin-wall regime and the same  arguments from Sec. \ref{sec:sbs_maxC} apply. For example, the estimate of the speed of sound, in analogy with the Eq. \eqref{eq:linEoS_cs_qb}, is
\begin{eqnarray}
(c^2_s)_a \approx \frac{\varphi_c (4 \varphi_c-3)}{12 \varphi_c^2-15 \varphi_c+4} \,. \label{eq:linEoS_cs_qb_v4}
\end{eqnarray}
From the above and the LinEoS results \eqref{eq:LinEos_C_w} we can predict $\mathcal{C}_{\rm LinEoS}=0.337$ for the compactness of the maximaum mass configuration with $\Lambda=0.186$ ($\varphi_c=1.053$), whose true value is $\mathcal{C}=0.331$.

\begin{figure}
\centering
\includegraphics[width=0.42\textwidth]{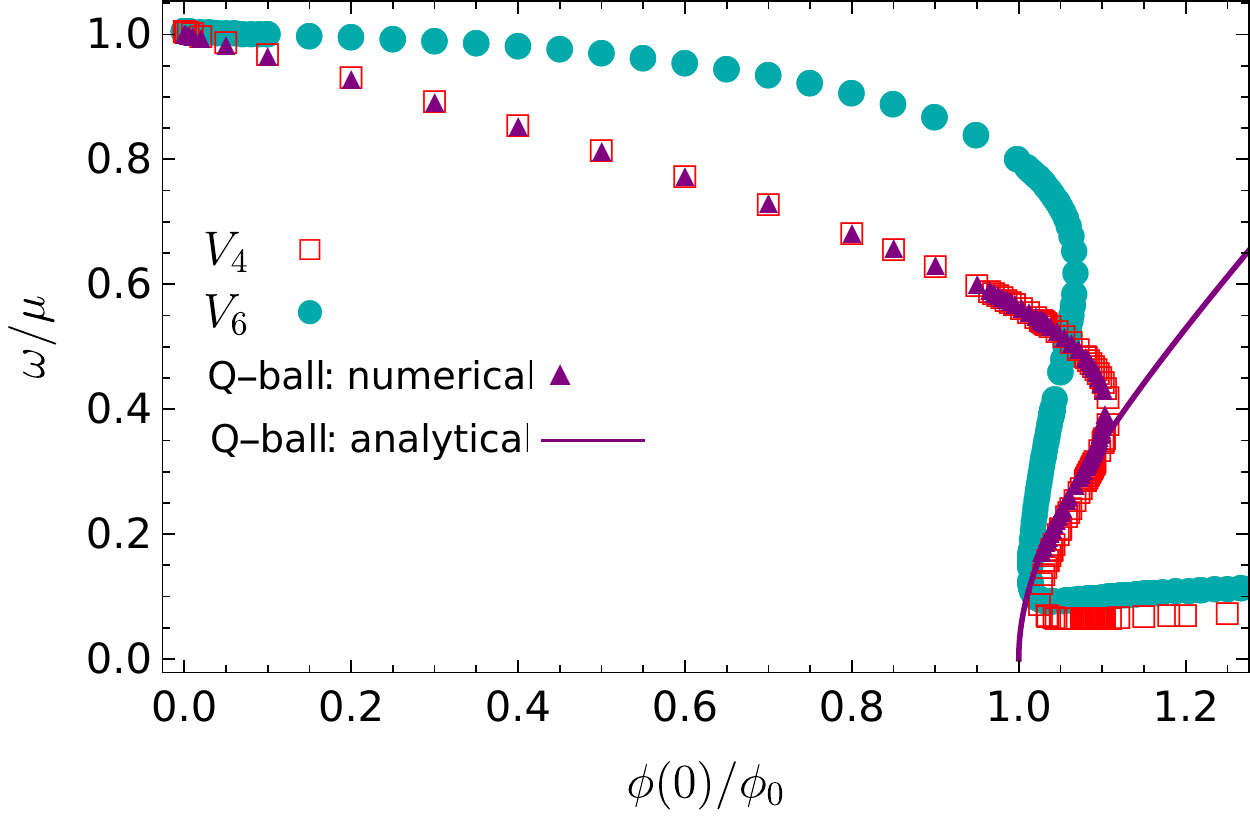}
\caption{SBSs scalar frequency as a function of the central field in the case of the quartic potential \eqref{eq:V4}, compared with the benchmark case \eqref{eq:v6_1}  for $\Lambda=0.186$. Q-ball (with the quartic potential) results, both numerical and analytical, are indicated with the purple triangles and the purple line, respectively.}
\label{fig:v4_fi_W}
\end{figure}

The presence of only one turning point in the $\Bar{M}-\varphi_c$ diagram [Fig. \ref{fig:v4_fi_M}] indicates that only one stable branch is inherited from flat spacetime, in contrast with the $V_6$ case. This branch is, as in the $V_6$ case, succeeded by the compact unstable branch, as elaborated using catastrophe theory arguments in \cite{Tamaki:2010zz}. The fact that both SBSs with a quartic potential and MBSs have only one stable and  one unstable branch, noted in \cite{Tamaki:2010zz}, is in fact accidental, as the MBS stable zone originates from the quantum pressure, while the stable zone of SBSs with the $V_4$ potential is inherited from the corresponding Q-balls.  This is the reason why the $\Bar{M}-\varphi_c$ relation in the low compactness stable branch for the sextic potential is described by the NBS scaling \eqref{eq:wf_mbs_fiw} in Fig. \ref{fig:v4_fi_M}, while the quartic is not. This difference can be also illustrated by the $w-\varphi$ diagram in Fig. \ref{fig:v4_fi_W}: for
the quartic potential
one can observe a much sharper decline of the $w(\varphi_c)$ curve from $w=1$ than with
the $V_6$ potential. Convergence between the two potentials occurs, manifest in both $\Bar{M}-\varphi_c$ and $w-\varphi_c$ representations (Figs. \ref{fig:v4_fi_M}, \ref{fig:v4_fi_W}), in the compact limit, where gravity does not discriminate between the highest powers of the scalar potential. In agreement with Sec. \ref{sec:sbs_an}, the  $\tilde{w}(\varphi_c)$ curve in the thin wall regime matches well the Q-ball result:
\begin{equation}
\tilde{w}=\sqrt{ 1-3 \varphi_c+2 \varphi_c^2} \,.
\end{equation}

\section{Conclusion}

In this work we have provided a comprehensive analysis of the self-gravitating solitonic objects made of complex scalars that obey potentials with
false/degenerate vacua.
We have built on previous studies by reinterpreting and improving them,
and also by providing novel results. In more detail,
we find that
in the thin wall regime, because of the presence of the non-trivial vacuum, these objects can achieve high compactness, saturating  the Buchdahl limit with the causality constraint $\mathcal{C}_{\rm B+C}=0.354$ (Sec. \ref{sec:sbs_maxC}). These values are
the highest
that have been found so far for motivated ECO models.  This results also provides a new perspective on SBSs - objects with an incredibly stiff equation of state, where the information on local disturbances is transmitted at almost the speed of light.

We have established the robustness of this picture by checking various potentials considered in the literature - general sextic, quartic and  cosine potentials (Sec. \ref{sec:par_space}). Although in this work we have stressed the general features of SBSs, particular models considered in the literature  present some differences. For the ease of navigating amongst different models, we summarise in Table \ref{tab:2} the potentials that we have considered in this work.   In addition, the bubble-like (in the thin-wall regime)  behaviour of the scalar field allows for an analytic description of these configurations, which we have presented for the simplest case of degenerate vacua (Sec. \ref{sec:sbs_an}), although we expect that this description can be extended (in a straightforward manner) to the other potentials considered in Sec. \ref{sec:par_space}. This analytic solution can be used as an approximation to the numerical one, and it has helped us obtain analytic control of  SBS solutions for arbitrarily small values of $\sigma_0/M_{\rm Pl}$.

\begin{table*}
 \begin{tabular}{||c c c c c||}
\hline
Boson star model & Potential & $M_{\rm Pl} \to \infty$   & $
|\Phi| \ll M_{\rm Pl} $  &  $\mathcal{C}_{\rm max}$  \\[0.5ex]
\hline\hline
\hline
Mini (MBS) & $\mu^2 |\Phi|^2$ & / & Newtonian (NBS) & 0.11 \\
\hline
Self-interacting (SIBS) & $\mu^2 |\Phi|^2 + |\lambda| |\Phi|^4 $ & /  & NBS: self-interacting & 0.16  \\
\hline
Soliton (SBS): simplest & $\mu^2 |\Phi|^2 \Big(1 -2 \frac{|\Phi|^2}{\sigma_0^2} \Big)^2$ & Q-ball: simplest & NBS/MBS  & $\leqsim 0.354$ \\
\hline
SBS: sextic & $\mu^2 |\Phi|^2 -|\beta| |\Phi|^4 + |\xi| |\Phi|^6$ & Q-ball: sextic & NBS/MBS & $\leqsim 0.354 -0.06 w_0^2$ \\
\hline
SBS: cosine [``axion BS''] & $|\alpha| \Big[1- \sqrt{1- |\beta| \sin^2 \Big(\frac{|\Phi|}{\sigma_0} \Big)} \Big]$  & Q-ball: cosine & NBS/MBS  & $\leqsim 0.354$ \\
\hline
SBS: quartic & $\mu^2 |\Phi|^2 -|g| (|\Phi|^2)^{3/2} + |\lambda| |\Phi|^4 $ & Q-ball: quartic  & Q-ball: quartic & $\leqsim 0.354-0.2 w^2_0$  \\
\hline
\end{tabular}  \caption{Models considered in this work, their potentials and main properties (including the non-gravitating limit $M_{\rm Pl} \to \infty$, the low compactness limit $
|\Phi| \ll M_{\rm Pl} $ and the maximal compactness  $\mathcal{C}_{\rm max}$). For the quartic SBS we have quoted $\mathcal{C}_{\rm max}$ for the general potential \eqref{eq:v4_2} and both general quartic and sextic potential expressions are valid for small $w_0$. Value for $\mathcal{C}_{\rm max}[\rm{SIBS}]$ is taken from \cite{AmaroSeoane:2010qx}.\label{tab:2}}
\end{table*}

In the low compactness limit, the configurations are stabilized either by gravity (thus behaving as MBS or NBS, cf. Sec. \ref{sec:par_dilute}), or by self-interactions (like in the case of the quartic potential, cf. Sec. \ref{sec:par_v4}). For field values close to the Planck scale, the compact stable branch shrinks, and only the low-compactness one, described by the MBS model, survives (cf. Sec. \ref{sec:par_large_lam}).

One follow up to this project
is to generalize our formalism to study
scalar-fermion  solitonic configurations, which could  form through different channels and possibly develop other interesting phenomenology \cite{Lee:1987rr,Bahcall:1989ff,Hong:2020est,Gross:2021qgx}. This work is currently in progress and we will present  it elsewhere \cite{FSS}. There are several other topics that are a natural continuation of this work and could be of interest: investigating the applicability of our results for  non-Abelian global or gauged gravitating Q-balls or higher-spin fields; comparing properties of (gravitating) Q-balls to configurations made of real bosons , motivated by axion or vector DM \cite{Hui:2016ltb,Hui:2021tkt}, with a false/degenerate vacuum in the potential; obtaining analytic control and the parameter space behaviour of the SBS quasi-normal modes \cite{Macedo:2013jja} and tidal Love numbers \cite{Cardoso:2017cfl,Sennett:2017etc};  probing the I-Love-Q and other universal relations \cite{Yagi:2013awa, Yagi:2016bkt}; describing  GW echoes \cite{Urbano:2018nrs}  etc.

From the astro-particle physics perspective, it is also important to understand what class of Q-ball production mechanisms can lead to a significant fraction of compact objects,
and how can the presence or absence of  SBS signatures in present and future GW detectors help us constrain their formation mechanism
and possibly gain information on the early universe.
On the phenomenological side, it is therefore imperative to systematically explore the behaviour of SBSs in binaries and their GW signatures \cite{Palenzuela:2017kcg}, also for  low-$\mathcal{C}$ configurations \cite{Chia:2020psj}. Work addressing the last topic is reported elsewhere \cite{Bezares:2022obu}.

\begin{acknowledgments}
We acknowledge financial support provided under the European Union's H2020 ERC Consolidator Grant ``GRavity from Astrophysical to Microscopic Scales'' grant agreement no. GRAMS-815673. This work was supported by the EU Horizon 2020 Research and Innovation Programme under the Marie Sklodowska-Curie Grant Agreement No. 101007855.
We would also like to thank Alfredo Urbano for an initial collaboration on these topics, Miguel Bezares for useful conversations on boson stars and for checking our numerical solutions with an independent evolution code and Paolo Pani and Steve Liebling for comments on the manuscript.
\end{acknowledgments}

\appendix

\section{Numerical solutions of Einstein-Klein-Gordon system} \label{AppEKG}

In the following, we present the technical details regarding the well-posedness and  numerical solution of the Einstein-Klein-Gordon system  given by Eqs. \eqref{eq:u_struc}, \eqref{eq:v_struc} and \eqref{eq:SBS_KG}, which we reproduce here for clarity:
\begin{eqnarray}
\frac{1}{r^2}\l(r\,e^{-u}\r)' -\frac{1}{r^2}&=& -\frac{1}{M^2_{\rm Pl}} \rho\,,\label{eq:u_struc_app}\\
e^{-u}\l(\frac{v'}{r}+\frac{1}{r^2}\r)-\frac{1}{r^2}&=&\frac{1}{M^2_{\rm Pl}} P_{\text{rad}}\,,\label{eq:v_struc_app}\\
\phi''+\l(\frac{2}{r}+\frac{v'-u'}{2}\r)\phi'&=&e^u\l({\frac{dV}{d|\Phi|^2}}-\omega^2e^{-v}\r)\phi.\label{eq:SBS_KG_app}
\end{eqnarray}

First, let us notice that the structure equations have poles at $r=0$, and we thus have to impose regularity there. In addition,
we require solutions to be asymptotically flat. The resulting boundary
eigenvalue problem uniquely determines the eigenvalue $\omega$. In fact, on the one hand, a local expansion of the fields  
around $r=0$ up to $\mathcal{O}(\epsilon^4)$ yields
\begin{eqnarray}
u(\epsilon) &\approx& 0+\frac{1}{6} \Lambda ^2 \epsilon^2\varphi_c^2 \left(1+ e^{-v_c} w ^2+\varphi_c^4-2\varphi_c^2\right)   \,, \label{eq:App_in_1}  \\
v(\epsilon) &\approx& v_c-\frac{1}{6} \Lambda ^2 \epsilon^2\varphi_c^2 \left(1-2 e^{-v_c} w ^2+\varphi_c^4-2\varphi_c^2\right)  \,, \label{eq:App_in_2} \\
\varphi(\epsilon) &\approx&  \varphi_c + \frac{1}{6} \epsilon^2\varphi_c \left(1-e^{-v_c} w^2+3\varphi_c^4-4\varphi_c^2\right)\,, \label{eq:App_in_3}
\end{eqnarray}
where we can set
$\tilde{v}_c=0$ and $w \to \tilde{w}$ with a rescaling of the time coordinate.
On the other hand, the leading order asymptotic behaviour, as discussed in Sec. \ref{sec:sbs_asymp}, is given by Eqs. \eqref{eq:u_asymp}, \eqref{eq:v_asymp}, \eqref{eq:asymp_LO}, which we reproduce here for the reader's convenience:
\begin{eqnarray}
u_> &=& -\log{\Big(1-\frac{2\Bar{M}}{\mathsf{r}} \Big)}\,, \label{eq:u_asymp_app} \\
\tilde{v}_> &=& \tilde{v}_\infty+\log{\Big(1-\frac{2\Bar{M}}{\mathsf{r}} \Big)}\, \label{eq:v_asymp_app} \\
\varphi_{\infty} &\simeq& \frac{\mathcal{A}}{\mathsf{r} ^{1+\beta_>}}e^{-\alpha_> \mathsf{r} }  \label{eq:asymp_LO_app} \,, \\
\alpha_> &=& \sqrt{ 1-e^{-\tilde{v}_\infty} \tilde{w}^2 } \,,\\ \beta_> &=& \frac{\Bar{M}}{\alpha_>}(1-2 e^{-\tilde{v}_\infty} \tilde{w}^2)  \,,
\end{eqnarray}
One can then match the numerical solution from the interior, obtained with the initial conditions determined by \eqref{eq:App_in_1}-\eqref{eq:App_in_3} at some small but non-zero radius $\epsilon$, with the asymptotic expansion at the infinity, at a finite but sufficiently large matching radius (``direct shooting''). This can be done by solving the four junction conditions %
\beq \label{eq:junc}
\Delta u |_\bullet =0 \,, \Delta v|_\bullet=0 \,, \Delta \varphi |_\bullet=0 \,, \Delta \varphi' |_\bullet=0 \,,
\eeq
where $\Delta x  \equiv x_> - x_<$ and $\mathsf{r}_\bullet$ is the matching radius, in the unknowns $\varphi_c$, $\tilde{w}$, $\mathcal{A}$ and $\tilde{v}_\infty$.
Numerical integrations were performed using  \texttt{Mathematica}'s \cite{Mathematica} default stiff solver. The stiffness of the system in the thin wall regime requires an extraordinary amount of fine tuning for the eigenvalues. Examples of the precision levels needed in order to produce compact configurations with a shooting method are given in \cite{Macedo:2013jja}.

Note that this procedure is applicable to SBSs, MBSs, to  the potentials considered in Sec. \ref{sec:par_space}, and by taking $v \to 0$, $u \to 0$ also to Q-balls. (The asymptotics of Q-balls are discussed in Sec. \ref{sec:qb_simp}). Note that the local expansion \eqref{eq:App_in_1}-\eqref{eq:App_in_3} is given for the SBS model and is different for the other potentials.
We have validated our results by successfully reproducing the MBS and SBS configurations of \cite{Macedo:2013jja} and by
verifying that the
static configurations
that we find do not change when used as initial data in the evolution code of \cite{Bezares:2022obu}. The radius of our solutions is found by inverting $\Bar{M}(\mathsf{R})=0.99 \Bar{M}$ (using bisection) and the Noether charge is calculated numerically by integrating Eq. \eqref{eq:SBS_charge_GR}.  An alternative method for numerical calculation  is outlined in \cite{Kleihaus:2005me,Siemonsen:2020hcg}.

\section{Definitions of boson star radius}  \label{app:rad}

\begin{figure}
\centering
\includegraphics[width=0.42\textwidth]{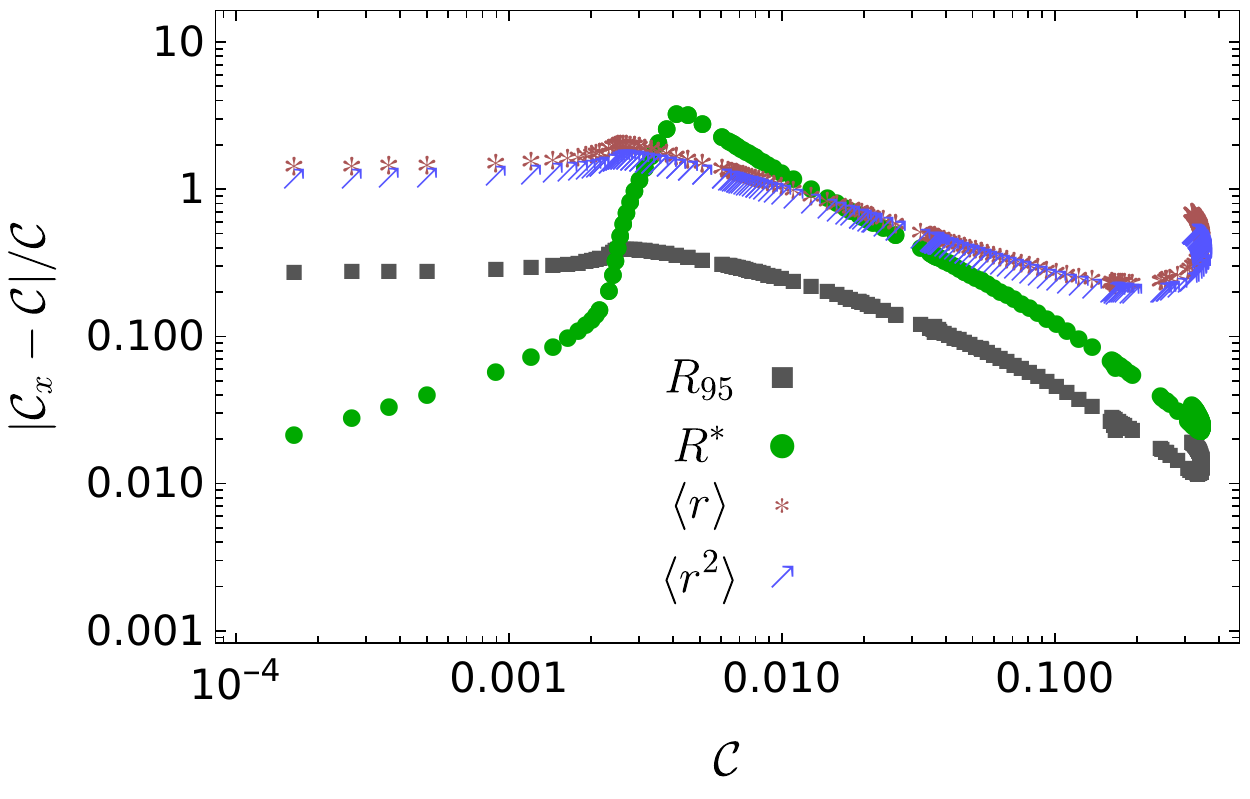}
\caption{Relative difference between $\mathcal{C}$ calculated with the benchmark definition of the radius enclosing 99\% of the mass and $\mathcal{C}_x=M/R_x$, where $R_x=\{R_{95}, R^\ast, \langle r \rangle,\sqrt{\langle r^2 \rangle}\}$.}
\label{fig:numerics}
\end{figure}

As the radius of boson stars is not well defined, various definitions have been used in the literature \cite{Schunck:2003kk}. Besides the definition used in this work (corresponding to 99\% enclosed mass),
the radius $R_{95}$ enclosing 95\% of the total mass  has also been used \cite{Barranco:2021auj}, as well as the following moments of the density distribution:
\begin{eqnarray}
\langle r \rangle = \frac{1}{M} \int^\infty_0 r \rho(r) dV \,, \label{eq:app_r_mass_av}\\
\sqrt{\langle r^2 \rangle} = \sqrt{\frac{1}{M} \int^\infty_0 r^2 \rho(r) dV}  \,. \label{eq:app_r_mass_av2}
\end{eqnarray}
Some authors have also considered radii enclosing a given fraction
(95\% or 99\%) of the total Noether charge $Q$ \cite{Palenzuela:2017kcg}, or the moments of $j^t$ \cite{Kleihaus:2011sx}. We will not discuss these definitions, because in this paper we are interested in the behaviour of Q-balls/SBSs as compact objects, for which  energy density based radii are more relevant. Finally, the inflection point $R^\ast$ was also taken as a Q-ball radius in \cite{Heeck:2020bau,Heeck:2021zvk}.

In Fig. \ref{fig:numerics}, we show the difference between our benchmark definition of compactness $\mathcal{C}=M/R$ and the compactness calculated with  (respectively) $R_{95}$ \,, $R^\ast$ \,, $\langle r \rangle$ \,,$\sqrt{\langle r^2 \rangle}$ for $\Lambda = 0.186$. The cutoff for the numerical integrals in Eqs. \eqref{eq:app_r_mass_av} and \eqref{eq:app_r_mass_av2} was taken to be the domain of the integration. As can be seen, while $R_{95}$ and $R^\ast$ produce  
relative differences in the value of the compactness in the compact stable branch of respectively $\sim 10^{-1} \,-\, 10^{-2}$ and $\sim 3 \cdot 10^{-1} \,-\, 2 \cdot 10^{-2}$  for $\mathcal{C} \geqsim 0.05$,  $\langle r \rangle$ and $\langle r^2 \rangle$ yield $\sim 0.2 \,-\, 0.7$ relative differences.
This difference
can be understood in the following way: in the flat space-time and $\omega \to 0$ limit, the scalar profile approaches a step function and $R^\ast$ approaches the hard surface radius. Neglecting the surface tension and the potential in that limit, one finds $\langle r \rangle \approx (3/4)R^\ast$ and $\sqrt{\langle r^2 \rangle} \approx \sqrt{(3/5)}R^\ast$. \\

\section{Analytic construction of Soliton Boson stars: technical details} \label{app:analytic_uS}

Here we provide some additional information on the analytic construction of SBS.

\subsection{$u_{B}$ metric coefficient on the boundary} \label{app:analytic_uB}

On the basis of the discussion from Sec. \ref{sec:ana_boundary} we find the $\ln(g_{tt})$ metric coefficient jump:
\begin{widetext}
\begin{eqnarray}
&& u_{B} =  - 2 \mathsf{m}_\ast  \mathsf{R}_\ast + v_< (\mathsf{R^\ast})- \log{\Big[\frac{\Lambda^2}{r} (e^{- 2 \mathsf{m}_\ast  \mathsf{R}_\ast + v_< (\mathsf{R^\ast})  } \{\chi_0+\chi_1 r +\chi_2 r^2+\chi_2 r^3\}+c_{\rm int}) \Big]} \label{app_eq_uB} \,, \\
&& \chi_0= \frac{1}{24} \left(-\frac{3 e^{-v_< (\mathsf{R^\ast})} w ^2 \text{Li}_3\left(-2 e^{2 \mathsf{m}_\ast  (r-\mathsf{R}_\ast)}\right)}{\mathsf{m}_\ast ^3}+\frac{3 \text{Li}_2\left(-2 e^{2 \mathsf{m}_\ast  (r-\mathsf{R}_\ast)}\right)}{\mathsf{m}_\ast ^3}+\frac{3 \log \left(2 e^{2 \mathsf{m}_\ast  (r-\mathsf{R}_\ast)}+1\right)}{\mathsf{m}_\ast ^3}\right) \nonumber \,,\\
&&  \chi_1= \frac{1}{24}  \left(\frac{24}{\Lambda ^2}+\frac{6 e^{-v_< (\mathsf{R^\ast})} w ^2 \text{Li}_2\left(-2 e^{2 \mathsf{m}_\ast  (r-\mathsf{R}_\ast)}\right)}{\mathsf{m}_\ast ^2}+\frac{6}{\mathsf{m}_\ast ^2 \left(2 e^{2 \mathsf{m}_\ast  (r-\mathsf{R}_\ast)}+1\right)}+\frac{6 \log \left(2 e^{2 \mathsf{m}_\ast  (r-\mathsf{R}_\ast)}+1\right)}{\mathsf{m}_\ast ^2}-\frac{6}{\mathsf{m}_\ast ^2}\right) \,, \nonumber \\
&&  \chi_2= \frac{1}{24}  \left(\frac{6 e^{-v_< (\mathsf{R^\ast})} w ^2 \log \left(2 e^{2 \mathsf{m}_\ast  (r-\mathsf{R}_\ast)}+1\right)}{\mathsf{m}_\ast }-\frac{6}{\mathsf{m}_\ast  \left(2 e^{2 \mathsf{m}_\ast  (r-\mathsf{R}_\ast)}+1\right)^2}+\frac{12}{2 \mathsf{m}_\ast  e^{2 \mathsf{m}_\ast  (r-\mathsf{R}_\ast)}+\mathsf{m}_\ast }-\frac{6}{\mathsf{m}_\ast }\right) \,, \nonumber \\
&&  \chi_3= -\frac{1}{6}  e^{-v_< (\mathsf{R^\ast})} w ^2 \,. \nonumber
\end{eqnarray}
\end{widetext}
The integration constant $c_{\rm int}$ is determined by matching the previous solution to  $u_<$ at $\mathsf{R_<}$ and $\text{Li}_n$ is the polylogarithm function.

\subsection{Details of the energy balance  calculation} \label{app:analytic_T_full}

Here we provide details on obtaining analytic  approximations in Sec. \ref{sec:en_balance_an}  . We will first approximate the complicated algebraic expressions  \eqref{eq:sbs_en_bal_2_a} - \eqref{eq:sbs_en_bal_2_d_s}. The sum $B_{\mathcal{E}}+C_{\mathcal{E}} + D_{\mathcal{E} B}$ is subleading, as the dominant contribution, proportional to $\mathsf{m}^2_\ast \Lambda^2 \mathsf{R^\ast}^2 $, cancels out. In the Q-ball limit, one also has $\mathsf{m}^2_\ast \sim 1$, which suppresses the term proportional to $(1-\mathsf{m}^2_\ast)$. In the more compact branch, we have numerically established that $(1-\mathsf{m}^2_\ast) \leqsim 2.5$; however, in that regime the terms $\sim 1/\mathsf{R^\ast}$ are subdominant relative to the volume originating terms in $D_{\mathcal{E} \, <}$, because $(\mathsf{R^\ast} \Lambda \tilde{w})^2 \sim 4$, while $1/\mathsf{R^\ast} \sim \Lambda \tilde{w}/2 \ll 1$ . Finally, the contributions with $w^2e^{-v_\ast}$ are suppressed, both due to $v_\ast>0$ and to the numerical pre-factors.

Leaving only the terms $A_{\mathcal{E}}+ D_{\mathcal{E} \, <}$ on the right hand side of  our  master equation Eq. \eqref{eq:sbs_en_bal}, approximating $\mathsf{m}_\ast \approx \exp{(u_<(\mathsf{R^\ast})/2)}$ and expanding in $\Lambda$, we the find Eq. \eqref{eq:sbs_w_T} which we reproduce here
\beq
\tilde{w} \approx \Lambda \label{eq:sbs_w_T_app} \frac{\sqrt{-\frac{T^4}{5}+6 T^2+36} \left(T^4+10 T^2+30\right)}{T \left(-T^4+30 T^2+180\right)}  \,.
\eeq
Inverting this equation, we find complicated expressions for the two branches, which can be approximated as
\begin{equation} \label{eq:sbs_T_approx_wC}
T\approx T(\tilde{w}_\cup / \Lambda) \pm \sqrt{3.1 \, (\tilde{w}-\tilde{w}_\cup)/\Lambda} \,, \tilde{w} \sim \tilde{w}_\cup
\end{equation}
where $T(\tilde{w}_\cup / \Lambda) \approx 1.6$. Using Eq. \eqref{eq:sbs_T_approx_wC}, the simple expectation for $v_\infty \approx v_<(\mathsf{R^\ast})+u_<(\mathsf{R^\ast})$ allows us to find an approximate behaviour for $w$, given in Eq. \eqref{eq:sbs_w_asymp_an}.

\subsection{On the errors of estimating the radius} \label{app:analytic_error}

In Section \ref{sec:an_finale},  the semi-analytic calculation develops a few percent systematic error near the maximum mass. There are two reasons for this. First, the expansion of $u$ and $v$ in the interior [expressions \eqref{eq:u_int}-\eqref{eq:v_int}] works well only for small $\tilde{w}$, i.e. for configurations similar to $\rm{II}$. In the thick-wall regime, the perturbative expansion outlined in Section  \ref{sec:inteior} does not work well by construction. Although the frequency $\Tilde{w}$ is higher for the configurations similar to III, the field derivative is still exponentially suppressed and physically these configurations have even thinner walls than the ones similar to II. Thus, one can expect that some different perturbative scheme could improve the analytic description of the metric coefficients for the configurations similar to III.

The second reason for the systematic error is that Eq. \eqref{eq:qb_lam}, which describes the deviation between $R$ and $R^\ast$,  does the best job for the configurations similar to II. For configurations similar to III, we have established numerically that the choice
\begin{equation} \label{eq:lam_app}
\lambda_{\rm III} = \frac{4}{2} \frac{2.66}{\delta} \,
\end{equation}
is the most appropriate one, while for configurations similar to I:
\begin{equation} \label{eq:lam_app2}
\lambda_{\rm I} = \frac{2}{2} \frac{2.66}{\delta} \,.
\end{equation}
This issue can be circumvented by performing the (cheap) numerical inversion  $4 \pi \int^{R}_0 dr \, r^2 \rho(r)  = 0.99 M  $ around the trial value determined by the semi-analytic algorithm, with a piecewise analytic approximation for $\rho$ (obtained from the field and the metric coefficients approximants). Finally,  the limits of the analytic description of the thin-wall Q-ball-like region in the parameter space (i.e. configurations similar to I) are discussed in \cite{Heeck:2020bau}.

Regarding the compact unstable branch: the numerical calculations indicate additional step-like features in the bulk of the scalar profile. This feature, not accounted for in our analytical description, is probably the origin of the increasing error in this branch.

\bibliography{references}
\end{document}